\documentclass[12pt]{article}
\usepackage{amsmath,amssymb,amscd,graphics}
\newcommand{\bs}[1]{\boldsymbol{#1}}

\begin{document}
\bibliographystyle{abbrv}

\begin{center}
{\bf \large Biological evolution through mutation, selection, and drift:}
\end{center}
\begin{center}
{\bf \large An introductory review}
\end{center}

\begin{center}
{\sc Ellen Baake$^1$ and Wilfried Gabriel$^2$}
\end{center}

\noindent {\em Zoologisches Institut,
    Universit\"at M\"unchen, Luisenstr.~14,
     D-80333 M\"unchen, Germany.}

\noindent  ${}^1${\em email: baake@zi.biologie.uni-muenchen.de}

\noindent ${}^2${\em email: wilfried.gabriel@lrz.uni-muenchen.de}

{\small
\tableofcontents
}

\section{Introduction}
Why a review on biological evolution in the Annual Review
of Computational Physics?
The questions raised are not, in the first place, physical ones,
and the results reviewed are only partly computational. However,
the past few years have seen a boost of activities in physics
directed towards biology, and expectations run high that
cooperation between biology and physics will constitute a
flourishing branch of science in the next century \cite{NatOp99}.

Physicists seem  particularly attracted by the field of
biological evolution. The topic lends itself readily to application
of concepts, analytical tools, and numerical methods from
statistical physics, with biological understanding profiting from
this new viewing angle. However, {\em population genetics}, which
describes biological evolution in mathematical terms, is a
venerable discipline, which was founded by Fisher, Haldane, and
Wright in the twenties (R.A.\ Fisher may be better known to
physicists as the father of modern statistics.). The field has
 developed to a high degree of sophistication
and is laid down in a far-spread-out literature, much of which
is hard to access even for biologists, and much more so
for physicists. Hence, there is a tendency
to run into avoidable pitfalls and to reinvent the wheel.
Even worse, well-established concepts are sometimes
misunderstood and redefined
in misleading ways, and redundant terminology and notation is being created
in the course of action. This considerably adds to  entropy, and
impedes communication.
The purpose of the present review is therefore twofold.
We shall first review some population genetic foundations in a nutshell.
Since the field is rather mature, many pointers will be set to
other reviews or even books.
We will then bring some
structure into the emerging field by bundling up recent work on
mutation-selection models. 
But first of all, we must define our subject more precisely.

`Evolution proceeds via mutation and selection.' In this
shortened version, the understanding of biological
evolution has entered everyday knowledge. Of course, it's
too short to be correct: actually, evolution consists of
a fair number of elementary processes, i.e.\
mutation, selection, recombination, migration, and drift
(i.e.\ fluctuations due to finite population size),
and it is not even decided whether mutation and selection
are the most important ones. They are, however,
weighty factors, and  may serve as a
case study of evolution as such. For the conceptual issues
related to the notion of evolution, we refer the reader
to Endler \cite{Endl86}.

 We shall restrict ourselves to
models of mutation and selection,
without and with genetic drift. That is, we must -- regrettingly
 -- exclude the fascinating topic of
recombination, and, with it,
 the evolution and maintenance of sexual reproduction.
Fortunately, there is a recent review available on this
topic \cite{FOC97}. But due to finite time,
space, and knowledge of the authors, the field must be narrowed
down even further. We shall concentrate on models which describe
{\em genetic variation  within single populations}. This excludes
macroevolution, speciation, and phylogeny (which are concerned
with variation {\em between} populations), as well as
adaptive walks
(which describe species as {\em genetically
homogeneous} entities). For the same reasons, models of
coevolution  will not be considered here. For a simple model
of interacting populations, we refer to Bagnoli and Bezzi (this volume).

In order to provide the uninitiated with some flavour of the
field, we will set out with a few relevant questions before embarking on the
basic models of population genetics. The latter describe the
evolution of the composition of a population under the
joint action of  mutation, selection, and drift,
 and will provide the foundations
on which to build. Since deleterious mutations and mutational degradation
have been major concerns in the context of asexual species,
we shall then dwell on the major phenomena
that have been considered in this context, namely
error thresholds, Muller's ratchet, and mutational meltdowns.
We shall finally turn to recent developments which, on the one
hand, show ways out of mutational degradation, and, on the other
hand, connect to dynamical
aspects of evolution.

We shall be mainly concerned with models aiming directly at the
genetic (as opposed to the phenotypic) level, thereby setting aside
much of quantitative genetics. Mutation-selection models with an
emphasis on quantitative genetics have been recently covered by a
review \cite{Bue98}, and a textbook \cite{LyWa98}.
 We strongly recommend the
simultaneous use of these and other reviews (which will
be mentioned while we proceed), as well as
the standard textbooks of population genetic theory.
Among them are the four volumes of Wright's `classic'
\cite{Wright68,Wright69,Wright77,Wright78},
Crow and Kimura (1970) \cite{CK70},  Ewens (1979)
\cite{Ewe79}, and Nagylaki \cite{Nagy92}. Molecular evolution
in particular is covered by Ratner et al. (1995) \cite{RZK+95}.
For the less
mathematically minded, Hartl and Clark (1997) \cite{HaCl97}
(student level book
with an  excellent bibliography), Li (1997) \cite{Li97} (comprehensive
overview of molecular evolution),
and Maynard Smith (1989) \cite{JMS89}
(for overwhelming biological intuition) are warmly recommended.

\section{Some questions}

Much attention has been devoted to questions related to the
equilibrium situation known as {\em mutation-selection balance}.
What is the equilibrium composition of the population under
the simultaneous action of mutation and selection?
How large is the {\em mutation load}, i.e.\ the
        unavoidable loss
        of fitness due to the production of less-well-adapted
        individuals by mutation (the `cost of variability', as it were)?
        How large is the genetic variability?
        Is mutation-selection balance sufficient
        to explain the observed variability?
        This comparison of data and theory has, in the past, led to crucial
        insights both in classical and molecular genetics,
        for review, see \cite{BaTu89,Gill91}.

Apart from such equilibrium considerations, dynamical aspects
have become more and more important.
Can  present-day genetic variation be used
         to draw conclusions about the (past) operation of selection?
        In particular, inference of selection at the molecular
        level is currently a `hot topic'
        (cf.\ \cite{Taji89, HKA87, FuLi93} for a few original
        landmark papers,
        \cite{Aqua97,Huds94} for reviews, and \cite[Ch.~9]{Li97}
        for a comprehensive overview.)

Often, the joint action of mutation, drift, and selection
leads to a steady state with a stationary phenotype under
an ongoing turnover of genotypes.
What are the rules that govern this
turnover?
The latter question is very important for between-population comparisons
and also touches on {\em molecular phylogeny},
the art of reconstructing evolutionary
trees from sequence data; for review, see \cite{SOWH95} and
\cite[Ch.~5 and 6]{Li97}.

\section{Mutation, selection, and drift}
\subsection{Mutation and selection: The basics}

{\bf The basic equations:}
Loosely speaking, a {\em gene} is a portion of the genome which codes
for something, for example an enzyme. When special reference is
made to the location (the {\em site}) in the genome, one speaks of
the gene {\em locus}. A gene may occur in several
versions called {\em alleles}.

Let us, for the moment, consider a population of haploid
organisms which carry but one set of genes or chromosomes per cell,
like viruses, bacteria, or blue-green algae,
 which reproduce asexually. We shall focus
on one single gene with $K$ alleles, $A_1, \ldots, A_K$,
which will sometimes be called `types' in what follows.
It is further assumed that fitness is  solely determined by the allele at the
locus in question, i.e.\ variability at all other loci will
be ignored. Individuals may then be characterized by,
and  identified with, alleles.

Let us first assume that generations are discrete and  nonoverlapping.
Individuals are counted at the beginning of every
new generation (`before selection') and undergo the simple
process
\begin{equation}
\begin{CD}
 A_i @>v_{ji} w_i >> A_j \,.
\end{CD}
\end{equation}
Every $A_i$ individual produces, on average, $w_i\geq 0$ offspring
for the next generation, and then dies; included in $w_i$
are the  probability of survival to
the reproductive age (the {\em viability}), and the number of offspring
(the {\em fecundity}).
At every  reproduction event, mutation may occur,
i.e.\ the offspring of $A_i$ is of type $A_j$ with probability $v_{ji}$.
The mutation matrix ${\cal V}=(v_{ij})_{1 \leq i,j \leq K}$
is a Markov matrix, i.e.\ $v_{ij} \geq 0$ and
$\sum_{i=1}^K v_{ij} = 1$. The quantities $w_i$ are known as the
{\em Wrightian fitness values}, cf.\ \cite{CK70}. If the average
number of offspring is the same for all genotypes and only
viability differences are considered, one speaks of {\em viability
fitness}. In both cases, ${\cal W}$, the diagonal matrix
$\text{diag} (w_1, \ldots, w_K)$,
may be understood as the reproduction matrix (one should not
use the term `fitness matrix' here, since this is reserved for
the fitness of diploid genotypes, as described below). We shall,
however, use the notion of {\em fitness landscape} for the
mapping from allelic space to fitness, due to its intuitive
appeal as a (high-dimensional) mountain range
with a population moving within
it like a cloud, `trying' to access the highest peaks.
The picture goes back to Wright (1932) \cite{Wright32}, who
coined the metaphor of an {\em adaptive
landscape}. However, two different meanings became associated with this
expression (cf.\ \cite{Gav97}), wherefore we prefer  the
term  `fitness landscape'.  Concrete fitness landscapes,
as well as mutation models, will be specified lateron.

\bigskip

Let us now consider a {\em population} of  individuals
which is so large that the frequencies of the various types
in the population may be treated
as continuous quantities,
and random fluctuations may be neglected. As long as there is
no restriction on population size, the change of the
composition of the population across generations is then
described by the linear difference equation
\begin{equation}\label{absdiscretemuse}
  x_i' = \sum_{j=1}^K v_{ij} w_j x_j\,, \quad \text{or} \quad \;
  \bs{x}' = {\cal V} {\cal W} \bs{x}\,.
\end{equation}
Here, $x_i$ and $x_i'$ denote the {\em absolute} frequencies of $A_i$
individuals in successive generations, and $\bs{x} :=
(x^{}_1, \ldots, x^{}_K)^T$, where $T$ denotes transpose.
The corresponding {\em relative} frequencies
($p_i := x^{}_i / \|\bs{x}\|_1$ where $\|\bs{x}\|_1 = \sum_i x^{}_i$ is the
total population size)  are, however, more interesting,
and more readily observable from population samples.
Application of (\ref{absdiscretemuse}) yields
the  {\em nonlinear} discrete dynamical system
\begin{equation}\label{discretemuse}
p_i' = \frac{\sum_j v_{ij} w_j p_j}{\sum_j w_j p_j} \,,
\end{equation}
where $ \sum_j w_j p_j =: \bar w$ is the {\em mean (Wrightian) fitness}
of the population. Note that $\sum_i p_i=1$, and, therefore,
a probabilistic interpretation is adequate, although the dynamics
itself is deterministic.

Eq.~(\ref{discretemuse})  is the haploid version of the
mutation-selection equation
of  population genetics  as originally formulated
by Haldane (1928) \cite{Hal28}, reviewed in detail by Crow and
Kimura (1970)  \cite{CK70},
and recently  by B\"urger (1998) \cite{Bue98}. We have derived it here
for the most basic situation: a haploid population,
discrete time, infinite population size (i.e.\ no genetic drift),
and unconstrained population growth. In what follows, these
assumptions will be relaxed one by one.

 \bigskip
{\bf Population dynamics:} The mutation-selection equation
(\ref{discretemuse}) may seem of
limited relevance due to its derivation from unconstrained population
growth, which is clearly unrealistic. However,
 its range of validity is much larger. Consider a
scenario where, in addition to genotype-specific reproduction,
some kind of population regulation  is in effect which
eliminates a fraction $g$ of individuals {\em regardless of their genotypes},
i.e.
\begin{equation}
 \begin{CD}
   \bigcirc @<g<< A_i @>v_{ji} w_i (1-g) >> A_j \,.
 \end{CD}
\end{equation}
Here, eliminated individuals are symbolized by a hole (`$\bigcirc$').
The fraction $g$ may vary from generation to generation; in particular,
it may depend on the current population size in a nonlinear
fashion. For example, $g$ may
constrain population size to a maximal carrying capacity,
as in the case of the well-known Verhulst or logistic equations
(see, e.g., \cite[Ch.~2]{Mur93}).
Although the dynamics of absolute frequencies may differ
drastically from that predicted by Eq.~(\ref{absdiscretemuse}),
the dynamics of {\em relative frequencies} is easily shown to be
again described by Eq.~(\ref{discretemuse}) for {\em arbitrary} $g$,
cf.\ \cite{CK70}.
This goes together with the fact that, in contrast to
Eq.~(\ref{absdiscretemuse}),
Eq.~(\ref{discretemuse}) is obviously invariant under the transformation
$w_i \rightarrow w_i \cdot c$ for any (positive) constant $c$, if it is
applied to all fitness values simultaneously. That is,
{\em ratios} of fitness values determine the dynamics, rather
than absolute values.

As long as populations of moderate size are considered, focusing on
relative frequencies seems appropriate. However, this is beside the
point when populations get close to extinction; then absolute
frequencies {\em must} be considered. We will come back to this
point later.

\bigskip
{\bf Basic properties of mutation-selection models:}
 The explicit solution of Eq.~(\ref{absdiscretemuse})
is, of course,
\begin{equation} \label{linsol}
  \bs{x}^{(n)}=({\cal V}{\cal W})^n\bs{x}^{(0)}\,,
\end{equation}
where we have used $n$
to indicate generation numbers. From this, the solution of
Eq.~(\ref{discretemuse}) is obtained by normalization.
Without mutation, i.e.\ $v_{ij}=\delta_{ij}$, Eq.~(\ref{discretemuse})
reduces to a haploid version of
{\em Fisher's selection equation}. The explicit
solution of this haploid version is trivial,
 since ${\cal W}$ is diagonal. It is well-known
that mean fitness acts as a Lyapunov function for the dynamics,
 i.e.\ it increases along all trajectories; see, e.g.,
\cite{Ewe79,HS88}.  This is intuitively obvious since
fit individuals  flourish at the expense of less fit
ones -- this is what is called {\em selection}. As
a consequence, only the fittest type(s) (of those initially
present) will survive in the long run.

If mutation is present, too,  the analytic solvability of the
dynamics depends on whether
${\cal VW}$ can be diagonalized
explicitly, which is rarely the case for large $K$.
The time evolution may still be determined numerically
by van Mises iteration (see, e.g., \cite[p.~178]{Zurm84}) of (\ref{linsol}).
If ${\cal V}$
is primitive,  which is usually the case in biologically
relevant situations, existence of and global
convergence towards
a stationary distribution is guaranteed by the Perron-Frobenius
theorem, cf.\ \cite[Appendix]{KT81}.
This stationary distribution is   given by the
Perron-Frobenius eigenvector of ${\cal VW}$ if supplied with positive
sign and correct normalization; cf.\ \cite{Thom74,Mor76}.
Global convergence also guarantees the stability of the
van Mises iteration or related numerical procedures.

\subsection{A few extensions} \label{ext}
{\bf Diploid models:} So far, we have only mentioned haploid
organisms, i.e.\ those with only
one set of genes per cell. In most `higher' organisms, however, two
copies are present, at least in certain stages of
 their life cycles; i.e.\ they are {\em diploid}.
If the phase of the life cycle on which selection acts is haploid
(as in algae and mosses), Eq.~(\ref{discretemuse}) holds without
modification (note that we always exclude recombination).
With most species, however,
fitness is a function of the diploid genotype.
 Let
$w_{ij}$ be the fitness of an $A_iA_j$ individual, where
$A_i$ ($A_j$) is the allele inherited from the father (mother).
Since for most genes (the {\em autosomal}, as opposed to
{\em sex-linked}, genes),
$A_iA_j$ is indistinguishable from $A_jA_i$, one also has
$w_{ij}=w_{ji}$. Eq.~(\ref{discretemuse}) must be modified
according to the mode of reproduction. If the diploid genome is
passed on to the offspring without reshuffling (as in vegetative
reproduction in plants), every genotype $A_iA_j$ may be considered
as an entity, and nothing but a simple relabelling is required. If,
on the other hand, reshuffling of alleles takes place via random
mating and sexual reproduction (note that this does not {\em per
se} imply recombination), alleles are combined independently, i.e.\
the frequency
 of $A_iA_j$ is $p_ip_j$ in the next generation;
this is the famous {\em Hardy-Weinberg
equilibrium}, cf.\ \cite{CK70,HS88}.
The dynamical equation for the allele frequencies
(\ref{discretemuse}) must then be modified by replacing the $w_i$
 by the corresponding {\em marginal fitnesses} $\bar w_i
:= \sum_j w_{ij} p_j$. In contrast to (\ref{discretemuse}), the
 resulting equations are inherently nonlinear. As a
consequence, there may be multiple steady states,
or   limit cycles (e.g.\ \cite{Aki79,Hof85,BaWi97}).
 Things become simple again, however,
if there is no {\em dominance}.
Absence of dominance
 means that
$w_{ij}=\sqrt{w_{ii} w_{jj}}$ for all $i$ and $j$, i.e.\ the
fitness of every heterozygote is the geometric mean of those of
the corresponding homozygotes\footnote{We have used the term {\em dominance}
in the sense of {\em dominance in fitness} here; this should not
be confused with {\em dominance at the phenotypic level}.}.
Then, the diploid equation
reduces to the original haploid one (\ref{absdiscretemuse})
 with $w_i := \sqrt{w_{ii}}$
for all $i$; cf.\ \cite[p.251]{HS88} and \cite{WBS95}.
 In order to avoid further complications,
 we shall,
in what follows,  adhere to this convenient special case,
although we are well aware of the fact that dominance  is
abundant in diploid organisms, and  may change
the evolutionary dynamics substantially.

\bigskip
{\bf Continuous time:} The literature on mutation-selection models
is partly in discrete, partly in continuous time.
We shall now briefly clarify the relationship between the various models.
There are actually {\em two} continuous-time versions, which correspond to
the processes
\begin{equation}\label{cdc}
\begin{array}{ccccc}
 \bigcirc & \stackrel{\, d_i}{\longleftarrow} &
       A_i & \stackrel{v_{ji} b_i}{\longrightarrow} &  A_i+A_j
\end{array}
\end{equation}
and
\begin{equation}\label{cdp}
\begin{array}{ccccc}
 \bigcirc & \stackrel{\, d_i}{\longleftarrow} &  A_i &
     \stackrel{b_i}{\longrightarrow} & 2 A_i \,. \\
     &            & \Big\downarrow\vcenter{
                   \rlap{$\scriptstyle{m_{ji}}$}}&& \\
     &            & A_j & &
\end{array}
\end{equation}
Here, the $b_i$ and $d_i$ are (instantaneous)
birth and death rates, respectively. We are considering
 approximations of (\ref{cdc}) and
(\ref{cdp}) which are continuous in  time
with at most one `event' occurring at any epoch; for example,
 multiple births may be mimicked by a larger
birth rate.

In scenario (\ref{cdc}),
mutation occurs exclusively on the occasion of reproduction
events, with probabilities $v_{ji}$, as in discrete time.
In contrast, mutation  is an independent
process in (\ref{cdp}) and occurs, at rates $m_{ji}$,
at any instant of the life cycle. The difference in notation
is to remind us of the fact that the $m_{ji}$ are transition {\em rates}
(and thus have the dimension $1/\text{time}$), with
$m_{ii} = - \sum_{j \neq i} m_{ji}$, whereas the $v_{ji}$ are transition
{\em probabilities} (which are dimensionless), and hence
$v_{ii}=1-\sum_{j \neq i} v_{ji}$.

As to the reproduction and death rates, $r_i := b_i - d_i$ are
the {\em Malthusian fitness} values. In scenario (\ref{cdc}), $d_i \equiv d$
 is assumed to keep things simple,
whereas this is not required for (\ref{cdp}).
With this in mind, the  differential equations corresponding to (\ref{cdc}) and
(\ref{cdp}) read
\begin{equation}\label{coupled}
 \dot p_i = \sum_j v_{ij} r_j p_j - \bar r p_i \,
\end{equation}
and
\begin{equation}\label{para}
  \dot p_i = (r_i - \bar r) p_i + \sum_j m_{ij} p_j \,,
\end{equation}
where $\bar r := \sum_j r_j p_j$ is the {\em mean (Malthusian)
fitness} of the population.
The first version is sometimes called the {\em coupled}
mutation-selection equation and was  studied by
Akin \cite{Aki79},  Hadeler \cite{Had81}, and others. The second version
is the {\em decoupled} or {\em parallel} version and seems to appear first
in \cite{CK70}. Biologically, they differ in the assumptions on
the mutation mechanism. The coupled equation describes mutation
as replication errors, whereas the parallel one understands them
as effects of radiation, free radicals or thermal
fluctuations. Which is the more relevant contribution in nature
is an issue under debate, but undecided. It comes down to the question
of whether the number of mutation events is
closer to constant in time or constant per generation (see the
discussion in, e.g., \cite{Gill87,Gill91} and \cite{Ohta93}).
These differences are important for the comparison of species with different
generation times, like mice and men. For our
within-population framework, however, the distinction
is not an important one. Typically,
 the two continuous as well as the discrete time
version give very similar results. For continuous time, this is so because
the parallel version emerges from the coupled one in the limit
of weak selection and mutation, as was shown by Hofbauer
(1985) \cite{Hof85}.

The comparison of discrete with continuous time requires a closer
look. Comparing the mutation and reproduction operators separately
gives
\begin{equation}\label{discrcont}
{\cal V}=\exp({\cal M}\tau) \quad \text{and} \quad
{\cal W}=\exp({\cal R}\tau),
\end{equation}
where ${\cal M}$ is the mutation matrix with
entries $m_{ij}$, ${\cal R}:= {\rm diag}(r_1, \ldots, r_K)$, and $\tau$ denotes
 the duration of one generation. In the limit $n \rightarrow \infty$
and $\tau \rightarrow 0$ with  $n \tau = t = \text{const}$,
it follows from the Trotter formula \cite{Reed1} that
\begin{equation}\label{shortgenlim}
  ({\cal VW})^n = \exp \Big ( \Big ({\cal M}\frac{t}{n} \Big )
\exp \Big ({\cal R}\frac{t}{n} \Big ) \Big )^n
 \quad \stackrel{n \rightarrow \infty}{\longrightarrow} \quad
  \exp(({\cal M}+{\cal R})t)\,;
\end{equation}
cf.\ \cite{WBG98}. Hence, the discrete time equation
converges to the parallel one in the
{\em short generation limit}.

At the same time,
(\ref{discrcont}) clarifies the relationship between the Malthusian
and Wrightian fitness concepts. Obviously, the invariance of the Wrightian
fitness under
$w_i \rightarrow w_i\cdot c$ translates into invariance under
$r_i\rightarrow r_i+c$ for Malthusian fitness, i.e.\
fitness {\em differences}  rather than ratios are relevant here
(this also entered the derivations of Eqs.~(\ref{coupled}) and
(\ref{para})). Likewise,
other relations change from multiplicative to additive,
e.g.\ the condition for absence of dominance now reads
$r_{ij}=(1/2) (r_{ii}+r_{jj})$ for all $i$ and $j$,
 where $r_{ij}$ is the Malthusian fitness
of genotype $A_iA_j$.

With this in mind, we may -- and shall --  move freely between the versions,
 using the most convenient one for each
question considered (or the one  authors
have happened to find themselves most familiar with).

\bigskip
{\bf Finite Populations:} Of course, all real populations are
{\em finite}, and this must be taken into account in many situations.
Let us therefore consider a haploid population of $N$
individuals, its size remaining
constant over generations. The number of offspring per
parent  will be a fluctuating quantity, however,
even in the absence of selection;
that is, the (absolute) frequencies $N_1, N_2, \ldots, N_K$ of
types $A_1, A_2, \ldots, A_K$
are {\em random variables}, subject to the constraint
$\sum_i N_i=N$.
Actually, the simplest (and most popular) picture is that
of parents  being sampled with replacement, each with
probability $w_i/(\sum_j w_j N_j)$, to give birth to a member
of the next generation, which may or may not be mutated.
For $w_i \equiv c$ (so-called {\em neutral evolution}), the number
of offspring per individual then follows a binomial distribution
with success probability  $1/N$, cf. \cite{HaCl97,DePe91}.
The resulting fluctuations of allele frequencies are known as
{\em genetic drift}.
In general (i.e.\ with mutation and selection),
 the transition from one generation to the next
is a Markov process defined by multinomial sampling, i.e.\
the transition probabilities read
\begin{equation}\label{wf_sampling}
  P(\bs{n'} | \bs{n}) = \frac{N!}{n_1'! \ldots n_K'!} \,
  \psi_1(\bs{n})^{n_1} \ldots
  \psi_n(\bs{n})^{n_K}\,,
\end{equation}
where $\bs{n}=(n_1, \ldots, n_K)^T$ and $\bs{n'}$ are the type counts
in successive generations, i.e.\ realizations
of the random variable $\bs{N} := (N_1, \ldots, N_K)^T$, and
\begin{equation}\label{wf_probs}
\psi_i(\bs{n}) := \frac{1}{N}
     \sum_j v_{ij} \frac{w_j}{\bar w} n_j
 \quad \text{with} \; \bar w := \frac{1}{N} \sum_k w_k n_k\,.
\end{equation}
This innocent-looking process is known as {\em Wright-Fisher sampling}
and describes the interplay of mutation, selection, and drift.
The sampling scheme (\ref{wf_sampling}) is quite
unwieldy, however, and is better studied through diffusion
approximations (at least as long as the number of dimensions is
small), see \cite{Ewe79}. An alternative is to start from a continuous-time
analogue of (\ref{wf_sampling}), i.e.\ a master equation,
known as the Moran model \cite{Mor58,Ewe79}.

In contrast to the situation with infinite population models,
we have, for the sampling process, assumed that some kind of population
regulation is in effect to keep population size constant.
 This goes together with the fact that, like the infinite-population
version (\ref{discretemuse}),
the Wright-Fisher model is invariant
under multiplication of the fitness values
with a constant.
As long as the reproductive capacity of a population (i.e.\
$N \bar w$) safely exceeds
the carrying capacity $C$ of the biotope,  this is reasonable.
If, however, for some reason or other, $N \bar w < C$,
absolute (as opposed to relative)
fitness values should be taken into
account, and population dynamics must be modelled explicitly.
After all, in this case the population may go to
extinction, and  then this is the one important thing to
consider (the gene frequencies in the dying population are
irrelevant). We note this here to be kept in mind
but defer treatment to later chapters.

\bigskip

{\bf Basic properties of the stochastic process:}
Quite generally, mutation increases genetic variation, whereas selection
and drift tend to reduce it. Understanding their simultaneous
action quantitatively, however, is tremendously difficult.
Below we list  a few limiting cases and corner stones; for review,
see \cite{Wright69}.

If mutation is absent (i.e.\ $v_{ij}=\delta_{ij}$),
 the genetically homogeneous states
($N_i=N$
for some $i$, $N_j=0$ for $j \neq i$) are absorbing, and every
population will finally end up in one of them, an event known as
{\em fixation}. Selection will show up through a higher fixation
probability for advantageous alleles. This is more pronounced
 if population size is large;
in small populations, random effects predominate. To be more precise:
when a new advantageous (or deleterious) mutant is introduced into
an otherwise homogeneous population,
its fixation probability is roughly $s$, provided $N$ is not too small
and $Ns>1$. Here,
$s$ is the selective advantage (i.e.\ the difference in (Malthusian)
fitness between mutant and wild type); otherwise,
the mutant allele behaves very similar to a selectively
neutral one\footnote{Let us
 mention that, for diploid organisms without dominance,
the situation is the same, but $N$ has to be replaced by $2N$.
If dominance is present, however, the results may change
considerably.}.
When more than two different fitness values come into play,
 the `trafficking of alleles poses
serious difficulties to analysis, and one has to rely
on simulations, cf.\ \cite{Gill94a}.
 We will meet concrete examples later.

If the mutation matrix ${\cal V}$ is primitive, the
process is strictly ergodic\footnote{Note that ergodicists call
{\em strictly ergodic} \cite{Mane87} what Markovianists call {\em ergodic}
\cite{KT81}.},
 and there is a stationary distribution, which is accessible for
a few simple selection schemes \cite{Wright69}.
Stationary distributions  may be understood as
time averages, which makes  sense if
$K$ is small and $N$ is large, so that all states are
visited in evolutionary (or Monte Carlo) time. When the number
of possible alleles becomes too large, as
is usual in the molecular context,
the population will  move through the allelic space as a (small)
cluster
which will `never' sample the space. Then, the infinite alleles
limit ($K \rightarrow \infty$) is appropriate \cite{Kim64,Ewe79}, and
other notions
of equilibrium must  be sought which refer to the genetic variability
within the cluster independently of its position. A lot is
known here
if selection is absent, i.e.\ if  mutation and
drift are the only evolutionary forces.
A large edifice of theory, Kimura's {\em neutral theory
of evolution}, has been developed on behalf of this case. It has
reached a high
degree of mathematical sophistication; for review see \cite{Kim83}
or the collection of Kimura's papers annotated by Takahata, \cite{Kim94}.
One important measure of within-population variability is
 the frequency distribution of alleles in a sample
without reference to their type; under neutrality, it is
 given by the celebrated
Ewens sampling formula \cite{Ewe72,Ewe79}. Parallels between
frequency distributions of population genetics and
frequency distributions of statistical physics
have been reviewed by Higgs \cite{Higgs95}.

\bigskip

\begin{figure}
\hspace{0cm}
\begin{center}
\scalebox{.75}{\includegraphics{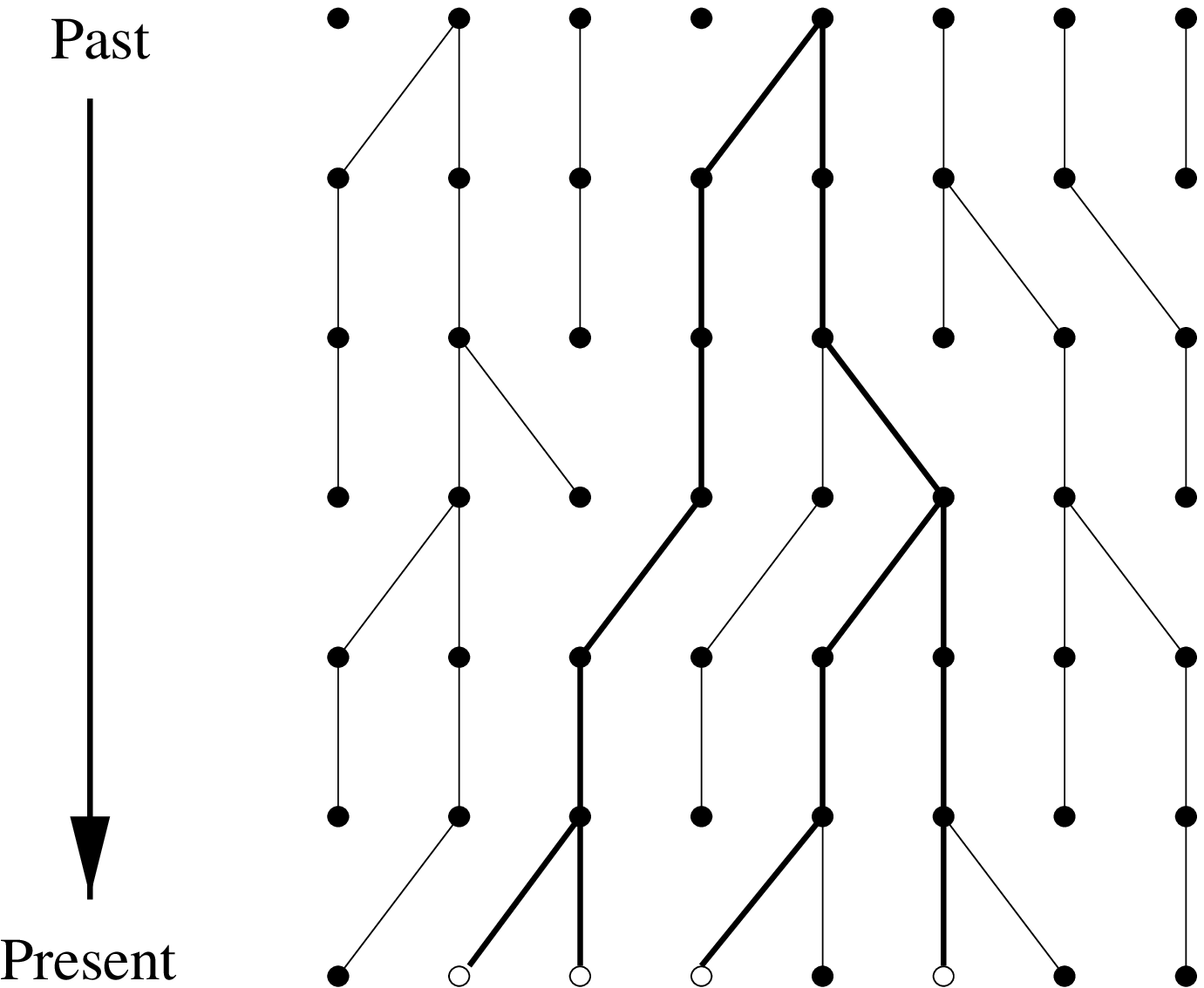}}
\end{center}
\caption{\label{sampling}}
\vspace{5cm}
\end{figure}

\begin{figure}
\hspace{0cm}
\begin{center}
\scalebox{1.}{\includegraphics{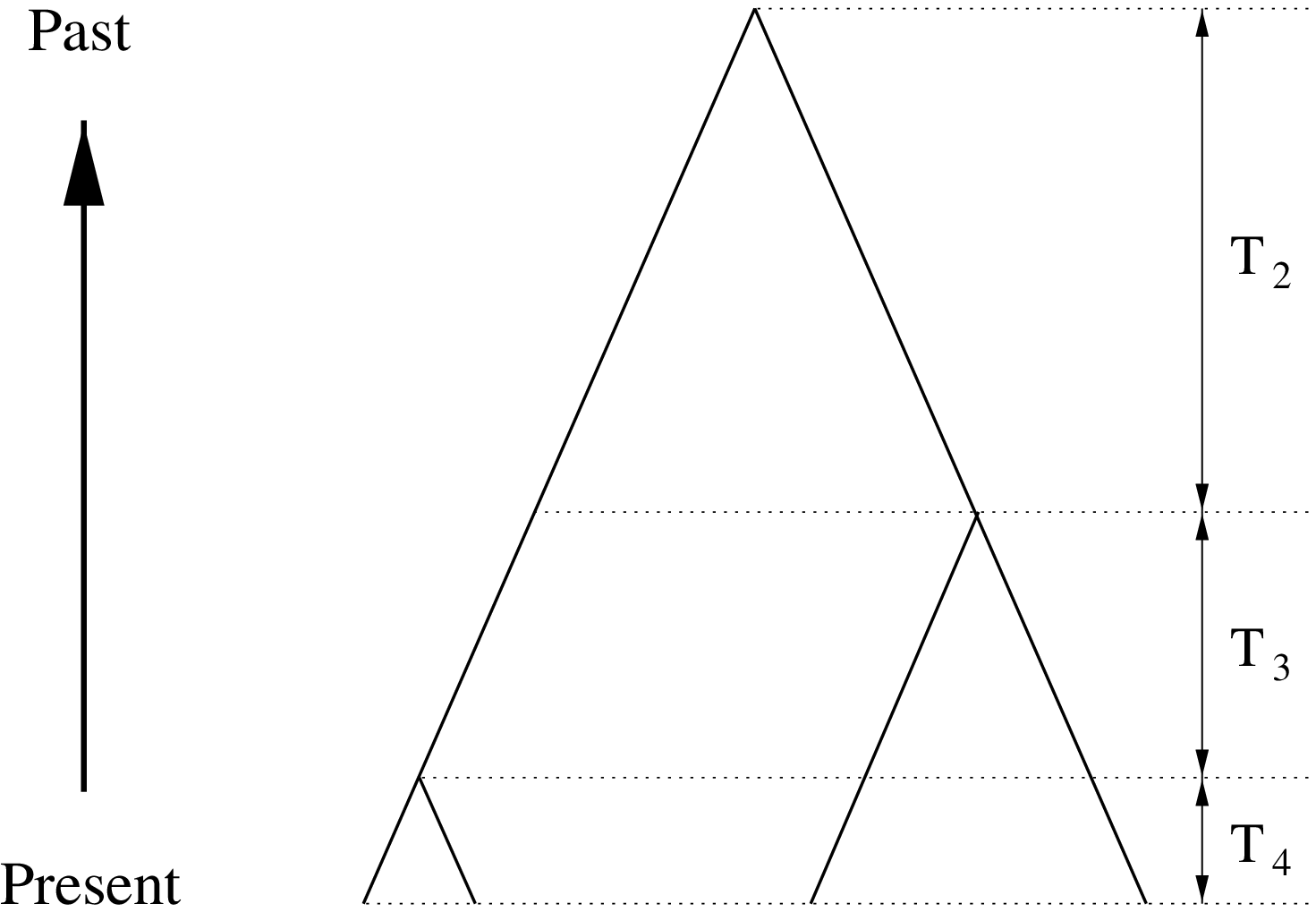}}
\end{center}
\caption{\label{coal}}
\end{figure}

{\bf Finite populations, backward in time:}
A major breakthrough which greatly advanced both theoretical
understanding as well as statistical inference from sequence samples
was achieved by Kingman (1982) \cite{King82a,King82b},
who considered the
evolution of finite populations backward in time.
The sampling process (\ref{wf_sampling}),
 which may be considered as a bifurcation process
forward in time, is then replaced by a {\em coalescent process}
backward in time; cf.\ Figs. \ref{sampling} and \ref{coal}.
The power of the backward process lies in the fact that only lineages
 surviving
to the present need be considered. For the purpose of inference,
these are further reduced to those  lineages leading up
to the sampled individuals. In any case,
quantities of prime interest are the statistical  properties
of the genealogies, i.e. the distribution
of coalescence times, and, in particular, of the time back to
the most recent common ancestor of a sample of individuals.
For neutral evolution,
coalescence theory is now well understood and has been
reviewed by Hudson \cite{Huds90}, and Donnelly and Tavar{\'{e}} \cite{DoTa95}.
The (relative) accessibility  is due to the
fact that mutations have no effect on the genealogies because,
by definition, neutral mutations do not affect the number of offspring
of individuals bearing these mutations. As a consequence,
individuals are independent and equivalent with respect to the
coalescence process, and the mutation process may be considered
separately from the genealogical process.

The basic quantity of the genealogical process is the probability
of a {\em coalescence
event} of two or more out of $m$ individuals, $m \ll N$, in a given generation.
Noting  that all $m$ individuals have distinct ancestors (in
the preceding generation) with probability $\prod_{i=1}^{m-1} (1-i/N)$,
the coalescence probability is
\begin{equation}\label{coal_prob}
  P(m) = 1 - \prod_{i=1}^{m-1} \big (1-\frac{i}{N} \big ) = \frac{1}{N}
\binom{m}{2} + O\big (\frac{1}{N^2} \big )\,.
\end{equation}
The approximation ignores the simultaneous coalescence of 
more than two lineages, which 
is
 reasonable since $m \ll N$. Any two of the present lineages are
equally likely to form the coalescing pair. Now,
the probability that $m$ individuals have $m$ distinct ancestors in
each of the preceding $n-1$ generations, and that a coalescence
event takes place in generation $n$ backward in time, is geometrically
distributed,
\begin{equation}
  Pr(T_m=n) = \big(1- P(m) \big )^{n-1} P(m) \,,
\end{equation}
with mean value $P(m) \simeq \frac{1}{N} \binom{m}{2} =: \lambda$.
Note that in the limit of continuous time, the geometric distribution
approaches the exponential distribution with density function
$\lambda \cdot \exp (\lambda t)$.
With this in mind, it may be shown that
 the time back to the most recent common ancestor of
a sample of size $M$, $T_{MRCA}:=\sum_{m=1}^{M} T_m$, is exponentially
distributed with mean $2N(1-1/M)$.

A neutral mutation process may be superimposed on the
coalescence process. Statistical properties of genealogies
may then be translated
into statistics of genetic similarity.
This was exploited by Derrida and Peliti (1991)
 \cite{DePe91}, who rediscovered the (neutral) genealogical process
and found it to be equivalent to the annealed random map model from
statistical physics \cite{DeBe88}.
 This equivalence allows the convenient calculation
of various measures of genetic structure.

As soon as selection comes into play, however, the independence
of individuals is lost. Dependence comes in through
fluctuations of the mean fitness of
the population, which involves {\em all} individuals.
 This was long thought  an insurmountable
obstacle to the treatment of
the coalescent process with selection,
but a breakthrough has been achieved recently by Neuhauser and Krone (1997)
\cite{KrNe96,NeKr97} in the framework
of interacting particle systems.

Apart from its theoretical importance, the coalescent process is
now an indispensable tool for the inference of evolutionary
history from sequence samples.
The most powerful methods for this purpose
 involve maximizing the likelihood of the observed sample configuration
over the set of parameters of the
model class considered. To be more precise, the likelihood of the
observed data for a given parameter set $\Theta$ (which may specify
the mutation process as well as the reproduction process)  may be
written as
\begin{equation}\label{lik}
  L(\Theta) = \sum_G Pr(D|G) Pr(G|\Theta),
\end{equation}
where  $Pr(G|\Theta)$ is the probability of the genealogy $G$
given the parameter $\Theta$, and $Pr(D|G)$ is the probability of
the data $D$ given the genealogy, see, e.g., \cite{KYF95}.
 Computation of the overall
likelihood along the lines of the coalescent is
far less costly than the corresponding simulations forward in time.
However, it still demands a summation over a huge number
of genealogies. Most of them are so implausible that they
contribute almost nothing to the likelihood. Therefore,
some kind of importance sampling is indispensable. Different approaches
are in use, see \cite{GrTa94}, \cite{KYF95}, and \cite{FKYB} for a review.
Still, this kind of inference problem remains demanding from
the computational point of view, even for neutral evolution.
If selection is considered, too, it becomes a real challenge.

\bigskip

\section{Specification of ingredients: Genotypes and phenotypes,
mutation models, and fitness landscapes}

So far, we have assigned fitnesses to genotypes, thus bypassing the
phenotype. Ideally, however, the sequence {\em genotype $\rightarrow$
phenotype $\rightarrow$ fitness} should be considered, and
we shall do so where possible. It will be apparent, though, that
often either the genotype or the phenotype will suffer some
neglect -- this is inevitable in view of the notorious
inaccessibility of the genotype-phenotype mapping. To be
more precise, a certain tradeoff will be observed. If
modelling is aimed  at the genotypic level, one may easily formulate
mutation models which are plausible in molecular terms; however,
the mapping from genotype to fitness is necessarily artificial.
If, on the other hand, the phenotype is in the centre of attention,
one has plausible fitness functions, but the mutation model
lacks a microscopic underpinning.

{\bf Genotypes:}
In the most straightforward picture, genotypes are
identified with linear arrangements of $L$ sites. Each site $i$
is equipped with a variable $\sigma_i$ which may take values from a
set $V_i$. This way, a configuration may be denoted by
$\bs{\sigma} \in V_1 \times \ldots \times V_L$. In the classical context,
`sites' are identified with `(gene) loci', and `variables' $\sigma_i$
 with `alleles'; then, $\bs{\sigma}$ is the configuration of a
so-called {\em multilocus system}.
The $V_i$ may be very `large' sets (comprising all possible
alleles at a locus), but sometimes the simple lumping into
{\em wildtype} ($+$) and {\em mutant} ($-$) alleles is sufficient, i.e.\
$V_i \equiv \{+,-\}$. The corresponding genotype space, $\{+,-\}^L$,
was introduced and visualized by  Wright (1932) \cite{Wright32}.

In the molecular context, $V_i \equiv V$
 may be the nucleotide alphabet $\{\text{A(denin), G(uanin),
C(ytosin), T(hymin)}\}$;
however, a binary alphabet ($V=\{0,1\}$ or $V=\{+,-\}$) is often
used instead, where the variables  are lumped into purins (A,G)
and pyrimidins (C,T). In both cases, configurations may be
interpreted as DNA sequences (or RNA sequences if the letter T
is substituted by U(racil)). Consequently, $V^{L}$
is known as {\em sequence space}.

While the classical picture was the primary one historically,
it may be considered as an {\em effective theory} today -- like
some kind of Landau-Ginzburg-Wilson theory in physics.
For the purpose of this review, we shall move freely
between the (molecular) sequence and the
(classical) multilocus pictures. In both cases, a continuous limit is
often appropriate, although genetic
information is discrete in principle.  This may be achieved through $L \rightarrow \infty$
(the {\em infinite sites model} \cite{Kim69a}, see also \cite{Ewe79}),
or by choosing $V$ as infinite or even continuous.

In certain classical contexts, the `genotype' is
an indirect construct. Rather than being
defined mechanistically as a sequence of letters,
it is defined through its effect on the phenotype, as will become clear
in a moment.

\bigskip

{\bf Phenotypes:} We will single out two phenotype spaces,
which are representatives of the molecular and the classical pictures,
respectively.

In the molecular picture, if taken seriously, it is clear right from the
beginning that the genotype-phenotype mapping is vastly complicated.
No matter what is considered as the phenotype, it will certainly
include several levels of organization, the first of
which is {\em protein folding} -- one of today's big unsolved problems.
One useful compromise is, therefore, to consider genes which code for
RNA (as opposed to proteins),  or   RNA molecules as such,
populations of which may be replicated in the lab \cite{Bieb87,BiGa97}.
Much of an RNA molecule's properties
is determined by its secondary structure, which, therefore,
serves as a legitimate phenotype.
Unlike DNA, which consists of two complementary, base-paired molecules,
RNA is single-stranded and partly folds back upon itself, forming
{\em stems} where base pairings occur, and {\em loops} where
letters are unpaired.
 The secondary structure is therefore
determined by the collection of base pairings
in a molecule, i.e.\ it is a {\em discrete} quantity.
It is computationally far more accessible than protein structure.
The prediction of the minimum free energy structure may be achieved
with the help of algorithms based on dynamic programming \cite{ZuSa84,HFS+94}.
A variety of  algorithms and different sets of thermodynamic
parameters have been used. Although the details of the results are
highly sensitive to the particular choices, several qualitative
features and many statistical properties (i.e.\ the frequency distribution
of structures) seem to be largely independent
of the prediction method, see \cite{TSBH+96} and references therein.
Exhaustive enumeration has been performed for all RNA
sequences with $L=30$ \cite{GGSR+96a,GGSR+96b}.
This reveals that the frequency
distribution of structures follows a generalized version of
Zipf's law \cite{GGSR+96a}.
With increasing length, there are only few common structures
and many rare ones, so that
only a few phenotypes will matter in practice. Sequences folding
into these common structures percolate sequence space.
On the other hand,
sequences folding into almost all common structures can be found
within a small distance of any random sequence \cite{GGSR+96b}.

Although it may be
debated whether the phenotype is  representative,
 it is felt that it captures some typical features of  biological
macromolecules, like long-range, asymmetric and irregular
interactions, and a many-to-one mapping from genotype to phenotype.

In the classical picture, the phenotype space is a {\em trait space}.
Most traits are so-called {\em quantitative traits}, i.e.\ they
either vary continuously (e.g.\ body height or milk yield), or
there is a large number of possible values that may be adopted
(e.g.\ the number of bristles on the abdomen of {\em Drosophila}).
This is opposed to
discrete traits (peas may be either green or yellow). For
practical reasons, one is often restricted to one or a few
 traits. The mapping from genotype to phenotype relies on the assumption
that a large number of loci contributes  to
 a given trait  $z$. In the simplest case, they act independently,
so that $z$ may be written as $z=\sum_j \alpha_j \sigma_j$,
 where $\alpha_j$ is known as the {\em effect} of site
  $j$ on the trait, and $\bs{\sigma} \in \{0,1\}^L$.
Similarly, $z_i=\sum_j \alpha_{ij} \sigma_j$ if $\bs{z} \in \Bbb R^n$.
Much of the quantitative genetics literature
assumes $\alpha_j \equiv \alpha$, but loci of major and minor effects
have been identified recently (for review, see \cite{TMO95}),
 which reveals the relevance of
 inhomogeneities across sites.
 However, even the homogeneous
models have been extremely successful in describing and predicting
breeding experiments, i.e.\ short-term (artificial) evolution.

\bigskip

{\bf Genotypes from phenotypes:} Owing to the ready observability
of the phenotype and the virtual inaccessibility of the genotype
in many situations,
the phenotype is often considered as the primary quantity, whereas the
genotype is a derived construct. A prominent example is
quantitative genetics. For a trait vector $\bs{z}$,
$\bs{x}$ denotes the corresponding genetic contribution.
In general, a trait
is determined by both genetic and environmental contributions.
In line with the fitness landscape picture, we ignore environmental effects
for the purpose of this review,
thus reducing quantitative genetics to the  case
$\bs{x}=\bs{z}$. We are well aware that this comes close
to castration of an important subject; however, this restricted view
 will suffice
to understand those mutation models and fitness landscapes
from quantitative genetics which we will meet  in the
context of mutational degradation.

\bigskip

We have, so far, defined genotypes and phenotypes. Building on this,
we shall now introduce mutation models (which act on the genotype)
and fitness landscapes (mappings from genotypes into fitness
values, ideally via the phenotype).
 In doing so, we shall
classify the material according to logical and/or mathematical
aspects. This way, models of molecular evolution may appear next
to those used in animal breeding. This should not be misunderstood as
neglect of the
historical context or the biological motivation; however,
it is felt that the classical and molecular fields should
(and do!) intermingle,
and much can be gained by considering their mutual relationships.

\bigskip

\subsection{Mutation models}
If the genotype is a collection of sites, it
 is usually assumed that all sites mutate independently
and experience the same transition probabilities. With
binary variables at the sites, mutation  is either
 chosen symmetric or unidirectional.
Symmetric mutation is more adequate for the molecular context,
whereas unidirectional mutation is often used in the classical
regime. The notion behind the latter is that,
actually, multiple alleles per site are assumed, but they are
lumped into a (small) `wildtype' and a (large) `mutant' class,
where mutations from wildtype to mutant are predominant
and back mutations negligible, due to sheer entropic reasons.

With $\bs{\sigma} \in \{+,-\}^L$ and symmetric mutation
with probability $p$ per
site at every reproduction event, the mutation probability from
$\bs{\sigma}$ to $\bs{\sigma'}$ reads
\begin{equation}\label{ss_mut}
  v_{\bs{\sigma'} \bs{\sigma}} = p^{d(\bs{\sigma'},\bs{\sigma})} (1-p)^{L-d(\bs{\sigma'},\bs{\sigma})}\,,
\end{equation}
where $d(\bs{\sigma'},\bs{\sigma})$ is the {\em Hamming distance} of sequences
$\bs{\sigma}$ and $\bs{\sigma'}$, i.e.\ the number of sites where $\bs{\sigma}$ and
$\bs{\sigma'}$ differ. We shall, in what follows, refer to
(\ref{ss_mut}) as {\em sequence space mutation}, since it is
mainly used in this context \cite{Eig71}.

Constant mutation probabilities over sites clearly represent an idealization
which is seldom realistic.
The existence of  {\em mutational hot spots}
is very well documented in molecular evolution,
for review see \cite{SOWH95} or \cite[pp.~74--77]{Li97}.
However,  many important
aspects are expected to be captured by the homogeneous model already.

Owing to the more abstract nature of the genotype, the mutation
 models of quantitative genetics are far less mechanistic.
For genotypes $\bs{x}$ from a (possibly continuous) state space $\bs{X}$,
let $f(\bs{x'},\bs{x})$ be the {\em mutation distribution},
i.e.\ the probability density for a
mutation from  $\bs{x}$ to $\bs{x'}$ conditional on the
 assumption of a mutational
event ($\int_{\bs{X}} f(\bs{x'},\bs{x}) d \bs{x'}=1$).
It is often assumed that $f(\bs{x'},\bs{x})=g(\bs{x'}-\bs{x})$;
this is a generalization of the {\em random-walk mutation model}
as introduced by Crow and Kimura \cite{CK64}.
One favourite choice for $g$ is a multivariate Gaussian
distribution, as used by  Lande (1976) \cite{Lan76}.
This is the only quantitative genetics mutation model which we
will need in what follows; we will therefore leave it at that
and refer the reader  to  textbooks and reviews,
 e.g.\ \cite{BaTu89,Bue98,LyWa98}, for more variety.

\bigskip

Quite generally,
the mutation model  is less of a worry than the fitness landscape.
The lack of knowledge about the latter is reflected
by the jungle of choices which are in use, and which we shall
now try and explore.

\bigskip

\subsection{Fitness \/ landscapes}
 We shall use the notion of fitness
landscape in the way brought up by Kauffman and Levin \cite{KL87}:
As a mapping
from genotype space into the  real numbers. We shall be exclusively
concerned with stagnant environments, i.e.\ the landscape is fixed.
This is a severe restriction and excludes, for example, extinction
events due to environmental catastrophes.
On the other hand, it is often legitimate for short-term evolution
and particularly so for evolution in the lab, where conditions
may be kept constant over a relevant time scale when organisms
with short generation times (like viruses) are used.

We shall proceed from the simple
to the more `complex'.
The trivial landscape is all {\em flat}.
It is the major theme of the aforementioned {\em neutral theory of
molecular evolution} developed by Kimura in the sixties
as a response to the fresh discovery of an overwhelming
and unexpected amount of variation at the molecular level; see \cite{Kim83}.
Since this was hard to explain under the
then-standard multiplicative fitness function (see Eq.~(\ref{mult}) below),
the bold consequence
was drawn that the vast majority of mutations is {\em selectively
neutral}, i.e.\ has no effect on fitness; this gave birth
to the neutral theory.

Although a flat fitness landscape does certainly not qualify
as a model for evolution in general,
the neutral model rightly serves as the basis
for analysis of molecular data from certain genomic regions
(like large parts of the mitochondrial DNA, which are
non-coding and may be taken to be free of selection in a good
approximation),
as an important null hypothesis (which is actually often
hard to reject!), and
as a toy model which allows explicit calculation of many
quantities of interest.

Some `local' information about the fitness landscape may be gained
from mutation accumulation experiments, where recombination is
prevented with the help of genetic tricks, and
selection is  relaxed as far as possible.
 In such experiments (performed
mainly with {\em Drosophila}), one measures   fitness (in terms of
viability, or number of offspring, or both) of
the progeny of {\em a single} genotype over many
generations, i.e.\ as a function
of the {\em average number of mutation events}. From the
 observed decrease in fitness, the deleterious mutation rate is
estimated and can be as high as one mutation per genome per generation
for detrimental effects  of a few
percent, and lethal mutations much rarer (see
\cite{Muk72,Ohni77,CroSi83,HHAC92,Rice94,CLL94,JoScho95,LyWa98},
\cite{Keig96,KeCa97} for a contrasting view, and \cite{Kond98} for a review).
Since the mutant genotypes are not accessible, the information
gained does not allow a reconstruction of the fitness landscape proper,
not even locally. The simplest landscape {\em compatible} with
the observations is one with a unique fittest genotype, the wildtype
$++++\ldots++$;  every `$-$' site corresponds to one deleterious
 mutation with detrimental effect $s$ each.
 If deleterious effects act independently across
sites, an individual
carrying $j$ mutations  has Wrightian fitness
\begin{equation}\label{mult}
w_j=(1-s)^j \,;
\end{equation}
this is the so-called {\em multiplicative fitness function}. The corresponding
Malthusian fitness is linear, $r_j=-\alpha j$,
where $\alpha=-\ln(1-s)>0$ in due course\footnote{Note that
Malthusian fitness does not suffer from being negative.};
this is often associated with a
{\em Mount Fujiyama landscape} \cite{Kauf93}.
In both cases, there is no interaction between sites  --
this is known as lack of {\em epistasis}.

These fitness functions are examples of
what we would like to call
{\em permutation invariant} landscapes, meaning that the fitness
of a configuration is invariant under permutation of sites.
If sites are equivalent for the mutation model as well,
a drastic reduction
of relevant dimensions  ensues. For $V=\{+,-\}$, to
which we will adhere in what follows, the number of `$-$'
sites in a configuration serves as a valid description
for most purposes (with the exception of some aspects of Muller's ratchet,
see below), and the dynamical equations simplify considerably.
The fitness optimum is  at the boundary of the (now one-dimensional) space;
one speaks of {\em directional selection}.

Epistasis, in its
simplest form, is inherent in quadratic fitness functions,
\begin{equation}\label{quadr}
r_j=-(\alpha j + (\gamma / L) j^2)\,,
\end{equation}
 cf.\ \cite{KiMa66,Charl90}.
If this function is monotonic with its
maximum at the boundary, one still has directional selection,
but, depending on whether fitness is a concave ($\gamma > 0$)  or
convex ($\gamma < 0$) function,
existing mutations have an aggravating or an alleviating effect
on further ones, which is termed {\em synergistic} (or {\em positive})
and {\em diminishing returns} (or {\em negative}) epistasis, respectively
\cite{KiMa66,Charl90}. The mutation accumulation
data seem to reveal some synergistic epistasis; see the discussion in
\cite{KiMa66}.

An extreme form of  epistasis (which changes from positive to
negative) is {\em truncation
selection},
where fitness is a step function of the number of deleterious mutations,
 i.e.\ for some $k$,
\begin{equation} \label{trunc}
w_j=\begin{cases}
     1 &  \text{for} \; j \leq k \\
     1-s & \text{for} \; j > k \,. \end{cases}
\end{equation}
Again, an extreme case of this is what was originally called
the {\em single-peaked landscape} (SPL).
Here, $k=0$, i.e.\ only one configuration in the space (the `wildtype')
has a selective advantage, whereas all others (the `mutants')
are equally unfit. Following recent usage, we rechristen it
{\em sharply-peaked landscape} in  agreement with the previous abbreviation.
The SPL was originally suggested as a model for
prebiotic evolution \cite{Eig71}. Even here, it should not be
considered as more than a toy model. It may well describe
certain restricted regions of the genome, for example
the few sites in the active centre of an enzyme, the function
of which is likely to be destroyed by almost any mutation
which comes along. On the other hand, a mutation which hits
the beta sheet in the body of a protein may go unnoticed --
this is an aspect of {\em neutrality}.
In general, it is
obvious that fitness landscapes should include
{\em compensatory mutations}, i.e.\ instances
where a mutation at a second site undoes the harm done by the first.
 In the SPL, there is
no chance for compensatory mutations,
no matter what the mutation mechanism. Nice examples of compensatory
mutations are present in the stem regions of RNA secondary structures.
Since these distinguish only between paired and
unpaired regions irrespective of the particular bases at the
individual positions,
base pairings may be destroyed by a point mutation,
but may be re-established by a second, compensatory one, cf.\
\cite{Ste96,Higgs98}. In general, the opportunity for compensatory mutations
depends on the mutation model as well as the fitness landscape. In
the multiplicative (\ref{mult}) and the quadratic (\ref{quadr})
landscapes, for example, there are no compensatory
 mutations if mutation is unidirectional,
but plenty of them when mutation is symmetric.

In the quadratic fitness function (\ref{quadr}), one may interpret $j$ as a
quantitative trait $z$. The
corresponding Wrightian fitness corresponds to a Gaussian distribution, which
may be written as a function of the trait value:
$w(z) = \exp(-s(z-z_{opt})^2)$,
 where $z_{opt}$ is the optimal
phenotype, and $s$ determines the strength of
selection\footnote{We use $s$ as a general variable to indicate
the strength of selection. It may express a selective advantage
or a disadvantage depending on the context.}.
If the optimum
is in the interior instead of at the boundary,
one has {\em stabilizing selection}. This is presumably
a common form of selection on quantitative characters
in the sense that selection favours intermediate trait values in preference
to either extreme. In contrast to the situation with directional selection,
the fittest genotype  is not unique in the configuration picture.
There is a large proportion of compensatory mutations,
which restore the fitness without restoring the original genotype.

\bigskip

We will next consider more elaborate landscapes and, for that purpose,
return to the  configuration space $\{+,-\}^L$.
Two questions will be crucial: How rugged is a landscape, and
how large is its degree of neutrality?

Neutrality is easily defined as the average number of neutral
neighbours per configuration, where `neighbourhood' is defined
with respect to the Hamming distance. Ruggedness is, intuitively,
related to the abundance of local peaks, as well as the depths of the
valleys separating them. Experimentally, the ruggedness of landscapes
has been explored in the multilocus context with the
help of special crossing experiments; for review see Whitlock et al.\ (1995)
\cite{WPMS95}. The authors  stress the abundance
of nonlinear interactions, with special emphasis on those
leading to multiple fitness peaks. A principal difficulty prevails,
however: Since it is impossible to explore all dimensions of
the genotype space, it is impossible to decide whether high spots
are isolated, or whether they are connected by ridges in higher
dimensions. One must therefore be satisfied to know that landscapes
are rugged, without knowing whether they are peaked. In line with this,
it has been argued on theoretical grounds that, in higher dimensions,
isolated fitness peaks are extremely
rare. High points are likely to be connected by a `bypass' in at least
one of the many dimensions \cite{Gav97,GaGr97}.

Formally, correlation functions and the density of local optima
have been proposed as measures of ruggedness; see, e.g.,
\cite{Kauf93,Stad96,ReSt98}.
 Since the exhaustive
exploration of a landscape is only possible if the number of sites
is small, one has to resort to some kind of statistical evaluation.
This may be done with the help of {\em adaptive
walks} \cite{Kauf93,MHP91}. An adaptive walk proceeds via random choice of
a neighbour with higher fitness in every step, until no higher
fitness is found; i.e.\ it stops at a local peak. The distribution
of the lengths of these walks, as well as the properties of endpoints
reached (e.g.\ whether they are isolated peaks or ridges),
 may be used to characterize a given landscape. Likewise,
{\em neutral walks} attempt to find a neutral neighbour in each
step so that the distance from the starting point increases \cite{GGSR+96a}.
The distribution of their lengths characterizes the neutral
properties of landscapes.

\bigskip

We shall now consider three representative
families of fitness landscapes in some more detail.

{\bf RNA, neutral networks, and holey landscapes}:
Since, within the RNA world, fitness is a function of the
(secondary) structure, an equivalent of our permutation-invariant landscapes
may be obtained by distinguishing one `target' structure
and assigning selective disadvantages to structures according to
some suitably defined distance from the target \cite{HSF96}.
 A corresponding SPL
then results from the assumption that all but the target structure
are equally unfit \cite{RFS98}.
However, assigning fitnesses to genotypes requires computation of the
structure, which involves an enormous computational effort.
Hence there is a large demand for
simpler toy models which mimick the essential features.
The pre-image of a given structure, its  {\em neutral space},
has been modelled as  the vertex set
of a certain random graph, with an edge between vertices if they are
neighbours in sequence space.
This way, one random graph (termed {\em neutral network})
 is associated with
every structure \cite{RSS97}.  A closely related concept is that of
{\em holey landscapes}, where
a fraction of randomly chosen genotypes is lethal \cite{Gav97,GaGr97}.
 The crucial parameter in both cases is the
mean fraction of neutral neighbours of a sequence. If it
surpasses a critical value, giant components of neutral networks, or
fit genotypes, appear
and percolate sequence space; this is as observed for the more common
RNA structures. The percolation
threshold decreases dramatically with the number of sites involved,
i.e.\ the dimensionality of the space \cite{RSS97,Gav97}.

\bigskip

{\bf NK landscapes, spin glasses, and Hopfield Hamiltonians:}
It is generally believed now that landscapes
are rugged due to the many interactions
between genes and within genes; but there is no agreement
how rugged exactly.  A family of {\em tunably rugged} landscapes
has been put forward by Kauffman and Levin \cite{KL87} and
further exploited by Kauffman and coworkers; for review, see
\cite{KaJo91,Kauf93,Kauf95}.
These so-called NK landscapes owe their name to the assumption that there
are $N$ (in our notation: $L$) sites, each of which interacts
with $K$ other sites in a random manner, as
originally motivated by metabolic \cite{Kauf69} gene regulatory
 \cite{Kauf86} networks. These landscapes correspond
 to spin glasses with $\tilde K=K+1$ interacting
spins, i.e.\
\begin{equation} \label{p_spin}
r_{\bs{\sigma}}=
   \sum_{\{i_1, \ldots, i_{\tilde K}\}} J_{i_1\ldots i_{\tilde K}}
\sigma_{i_1}\ldots \sigma_{i_{\tilde K}}\,,
\end{equation}
where the $J_{i_1\ldots i_{\tilde K}}$ are independent,
 identically distributed (i.i.d.)
random variables for each different set of indices
${\{i_1, \ldots, i_{\tilde K}\}}$. They can take positive as well as
negative values, thus causing  {\em frustration}.
The landscape is tunably rugged through $\tilde K$.
For $\tilde K=1$, one has an
(inhomogeneous) Fujiyama model, i.e.\ one unique peak and a highly
correlated landscape; it is also known as {\em random field
paramagnet} \cite{PBS97}.  For $\tilde K=2$, one has
the well-known Sherrington-Kirkpatrick
spinglass (for review, see \cite{MPV87});
it was  first suggested
in the evolutionary context by Anderson \cite{And83}.
It is rather
correlated but has a relatively large number  of local optima.
One special choice for the interaction constants is
$J_{ij}=\sum_{p=1}^{P}\xi_i^p \xi_j^p$. The resulting energy
function is known as {\em Hopfield's Hamiltonian} and was
originally designed to describe the operation of {\em neural}
(as opposed to {\em neuTral}!) {\em networks}.
Here, the
 $\bs{\xi}^p := (\xi_1^p, \ldots, \xi_L^p), \,p=1,\ldots,P$,
 are configurations
 randomly chosen from $V^L$. They represent the $P$
{\em patterns} learnt by, and stored in, a neural  network.

Hopfields's Hamiltonian
was used as a
fitness landscape by Leuth\"ausser \cite{Leut87a}
and Tarazona \cite{Tara92}. It may be rewritten as
\begin{equation}\label{hopf}
   r_{\bs{\sigma}} =  \frac{1}{L} \sum_{p=1}^{P}
   \big (\sum_j  \xi^p_j \sigma_j \big )^2\,.
\end{equation}
Here the ruggedness is determined
by $P$, in a statistical sense; of course it also depends on the
particular choice of the $\bs{\xi}^p$ in every single case.
For $P=1$,  one has
the so-called {\em Mattis Hamiltonian}. Noting that $\bs{\xi}^1$
may be chosen as `$+++++\ldots+$' without loss of generality,
it is clear that the Mattis Hamiltonian
corresponds to the permutation-invariant
quadratic fitness function (\ref{quadr})
with $\alpha=0$ and $\gamma=1$.

For $\tilde K=N$, one has what is called the random energy model \cite{Derr81}
in the terminology of spin glasses,
i.e.\ a fully uncorrelated landscape with a random
fitness value for every genotype. However, as a fitness landscape,
it is  older and has its roots in classical population genetics.
When thinking about the effects of mutation, Kingman (1977) stated
`the tendency for most
mutations to be selectively disadvantageous, presumably because they
upset the evolutionary house of cards, built up by the careful
improvement of many generations  $\ldots$. This suggests a model
$\ldots$
in which mutation always results in a completely novel allele,
and in which the fitness of the mutant is chosen from a fixed fitness
distribution' \cite{King77}. This became known as the {\em house-of-cards
model}.
In this formulation, the model is annealed whereas evolution on proper fitness
landscapes is a quenched problem, i.e.\ the interaction constants are
frozen during the evolutionary process. However, for reasons of
storage economy, the random energy model
is usually simulated as annealed in the evolutionary context
(e.g.\ \cite{APS89}), which
does not seem to produce noticeable artifacts.

For $\tilde K$ increasing from $2$ to $N$, there is an increasing
degree of conflicting constraints, as might be typical of
biological evolution, and `frustration' becomes more and more abundant.
The NK model has been extensively studied in the context of adaptive walks;
for review, see \cite{Kauf93}.

The hierarchy of spin glass models seems straightforward and
plausible. It should be noted, however, that the present state of
knowledge is fairly incomplete. Little is known rigorously, not
even about the energy functions as such (cf.\ \cite{MPV87}), let
alone the evolutionary dynamics on these landscapes.

One toy model which is between the smooth quadratic landscapes
and the rugged spin glasses is {\em Onsager's landscape} \cite{BBW97}
which is related to (\ref{p_spin}) through $\tilde K=2$ and
 $J_{ij}=\delta_{j,i+1}$.
Here, fitness is determined by the number of {\em domain walls}
in a sequence. As a consequence, there are compensatory mutations,
flat ridges, as well as a high degree of neutrality.
Of course, the symmetries are artificial, but, as with two-dimensional
Ising
models with nearest neighbour interaction, they make an exact solution
possible (see Sect.~5.1.2 below).

\bigskip

{\bf Multiple quantitative traits:}
Spin glass models do not have, and do not suggest,
 a phenotype, which is unsatisfactory
from the biological point of view. This is  different
for the  multiple quantitative traits (MQT) model which has been
put forward very recently  \cite{Higgs99}.
There are $L$ sites and $T$ traits, with every trait
influenced by $K \leq L$ randomly chosen sites.
For sequences as bit strings, $\bs{\sigma} \in \{0,1\}^L$,
traits are modelled as 
$z_i = \sum_{j=1}^L a_{ij} \sigma_j$, $i=1, \ldots, T$, with
$a_{ij}=1$ if site $j$ affects trait
$i$, and $a_{ij}=0$ otherwise. Wrightian fitness is based on
the sum of quadratic deviations of the traits  from their
randomly chosen optimal values, i.e.\
\begin{equation}\label{mqt}
w(\bs{z})=\exp(-s \sum_{i=1}^T (z_i - z_i^{opt})^2),
\end{equation}
where $s$ again measures the strength of selection.

The resulting fitness landscape  is tunably rugged
through the product $KT$. As long as $KT \ll L$, the sets of loci
linked to the different traits rarely overlap;
as a consequence, traits evolve independently, and
genotypes are possible which simultaneously optimize all traits.
If $KT \gg L$, on the other hand, most loci will affect more than one
trait, and there is no `perfect' genotype \cite{Higgs99}.

 The main difference between
MQT  and  spin glass models is that, in MQT landscapes, there are
clusters of high-fitness genotypes instead of maxima randomly
distributed throughout the space. This may be seen as a  higher-dimensional
issue of directional versus stabilizing selection. In
the former case, maxima are unique, and the degree of
neutrality increases with distance from them; in the latter
case, maxima tend to be plateaus with some degree of neutrality.

\bigskip

We have, so far, met a whole zoo of fitness landscapes, which
would benefit from a general characterization and classification.
An attempt in this
direction has been made recently through the concept of
{\em additive random landscapes} \cite{ReSt98,StHa99}. This is
a  large class of fitness landscapes both tunably rugged
and tunably neutral, which includes spin glass models, among
others.

\bigskip

\subsection{Scaling and limits}
 A very important issue for both mutation models
and fitness landscapes is that of {\em scaling} as a function of
the system size. For both of them,
extensive and  intensive scalings are in use.
{\em Extensive}
quantities scale linearly with $L$, whereas {\em intensive} ones are
independent of the number of sites. For the mutation
model, the distinction is straightforward on biological grounds.
 Extensive scaling is assumed with sequence space mutation
 (\ref{ss_mut}), and
motivated by the molecular picture: Every site has a fixed probability
to mutate, and thus the mutation probability for the whole
genome scales with genome size. In other cases,
 the genome (or the sequence, or the collection of sites)
is taken as the primary quantity, and with it associated a fixed mutation
rate $U$ per year or generation, which corresponds to
intensive scaling. In this context, one relies on the
{\em infinite sites limit}, i.e.\ $L \rightarrow \infty$ under
 $L \cdot p=U \equiv \text{const}$. Often, however, population
size must be considered, too. In what follows,
distinguishing carefully between various
limits and scalings will be crucial to an understanding of
apparently contradictory results. In particular, we shall
meet the
\begin{equation} \label{infgenlim}
 \text{\em{infinite genome limit:}} \quad L \rightarrow \infty \quad
 \text{under} \;  L \cdot p=\text{const}, \quad \text{followed by}
 \; N \rightarrow \infty\,,
\end{equation}
and the
\begin{equation} \label{infpoplim}
 \text{{\em infinite population limit:}} \quad N \rightarrow \infty \quad
 \text{followed by}
 \; L \rightarrow \infty\, \quad \text{under} \; p=\text{const}.
\end{equation}

Note that we adhere to the term  {\em infinite sites limit} in the above
sense if nothing is implied about population size.

Quite generally, for the real situation with finite $N$ and
finite $L$ but $N$ smaller than the number of possible configurations,
the population will move as a (relatively small) cluster in
the configuration space and eventually sample all of space,
but this will take extremely long. This is the motivation for the
infinite genome limit. A side effect is, however, that every site
experiences at most one mutation event ever. For example, if $++++++$
has mutated to $+-++++$ in one individual and to $+++-++$ in another,
 the configuration
$+-+-++$
can {\em never} be accessed in {\em any other}
 individual. Consequently,
the system cannot be ergodic, and the initial state
 of a population is important
even for infinite times. Deviations from such a model are apparent
in human mitochondrial sequence data, e.g.\ \cite{Weiss98}.

The infinite population limit, on the other hand,
 relies on the ergodicity of the
real system. It corresponds to the thermodynamic limit
in statistical physics, with $L$  the number of particles and
$N$ the number of copies of the system. The calculated quantities
are to be understood in the sense of time averages. It is
particularly adequate when the effective configuration
space is drastically reduced, e.g.\ due to an abundance
of lethal mutations. The
side effect here is that the averaging may take longer than
evolutionary (or Monte Carlo) time.

In summary, both limits may have their problems, which should be kept
in mind when interpreting the results. In particular, results obtained
with one of them may not be used to draw conclusions about the
other; this seems to have been overlooked in places. For example,
Derrida and Peliti \cite{DePe91} argue on the grounds of (\ref{infgenlim})
about the quasispecies
model (to be described in Sect.~5.1 below), which relies
on the limit (\ref{infpoplim}). This leads to results
at variance with the properties of the original model.

Let us now turn to scaling of fitness landscapes. For Malthusian fitness,
extensive scaling implies that
 $r_{max}-r_{min}$ increases linearly
with $L$. For Wrightian fitness, this corresponds to
$w_{max}/w_{min} \rightarrow \infty$ and, with the normalization
$w_{max}=1$, to $w_{min} \rightarrow 0$ for $L \rightarrow \infty$,
that is, genotypes become more and more lethal. With intensive
scaling, on the other hand, $w_{min}$ is bounded away from zero,
and there are no lethal genotypes.

 The SPL (sharply-peaked landscape) with $s < 1$ is an example of
intensively-scaled landscape (see, however, \cite{FrPe97}),
 whereas the multiplicative fitness function is extensive.

Both types of scaling are
used in both the molecular and the classical contexts, often
without awareness of the distinction. In particular, all
combinations of mutation and fitness scalings show up: all-extensive,
all-intensive, and mixed. It should be noted that the all-extensive
and the all-intensive scalings are related to each other
through a corresponding scaling of  (continuous) time, i.e.
$dt$ is replaced by $L dt$, see the discussion in \cite{WBG98}.
This is often overlooked
since only few investigations consider dynamical aspects at
all. In any case, it should be kept in mind that any real
 sequence is finite, and $L \rightarrow \infty$ is just a way
 of making life simpler --- a good one though,
since in these applications, 100 is often closer to
infinity than to one. When interpreting the results, attention
should be paid to correct adjustment of parameters to meet the
finite size context.

\section{Modelling mutational degeneration}
We have, so far, worked hard on the list of ingredients. As a reward,
 we are now in a position
to piece together mutation-selection models from the elements of
our construction set.

A large body of work on such models has actually been
related to mutational degradation in asexual populations.
These models predict upper limits for the mutation
rates above which mutation can no longer be controlled by
selection, the most important phenomena being error thresholds,
Muller's ratchet, and mutational meltdowns.

At first sight, the kind of question seems ill-posed. Wouldn't the
primary effect of mutation be to introduce new variation on which
selection can act to bring about evolutionary progress?
According to a commonly-held view, however, today's populations have
arrived at a state of elaboration where most mutations are
deleterious. Whereas this is a `finalistic' point of view which
one need not necessarily share, the argument is indisputable  for
two classes of problems:

1) Consider a sexual population
which, with the help of the larger adaptation potential of
recombination, has arrived at a fitness peak. Let then
a clonal lineage split off. Here, {\em clonal} means that
the genome is passed on unaltered
 from parent to offspring (apart from mutations).
This may occur through parthenogenesis
(in animals) or vegetative reproduction (in plants). Actually
clonal lineages arise frequently in both the animal and the plant kingdom
\cite{LyGa90}, even in fairly `high' organisms (e.g.\ certain lizards).

 In the absence of recombination,
such populations will incessantly slide off the fitness peak due to
deleterious mutations and genetic drift. This process is known as
{\em Muller's ratchet}. It may lead to the extinction of
populations through so-called {\em mutational meltdowns}
which provide one favoured explanation
for the observed short life of clonal lineages \cite{LyGa90}.

2) In the context of prebiotic evolution, and the evolution of viruses and
bacteria which do not have recombination (however, some viruses and bacteria
do recombine, see \cite{StHo87,JMS93,JMS94}),
 the mutation-selection equilibrium
is of primary importance.
With respect to the highest peak in
the landscape, every mutation is either neutral or deleterious.
Their joint effect on the equilibrium
fitness is known as the {\em mutation load},
$\ell := r_{\rm max} - \bar r$ (recall that $\bar r$ is the mean
(Malthusian) fitness of the population). This is important in the context
of the evolution of mutation (as well as recombination) rates.
After all, the evolution of repair mechanisms along with
that of the eukaryotic cell suggests that high mutation
rates are not unconditionally desirable. Actually, certain
models predict a complete loss of genetic structure when
the mutation rate surpasses a critical value; this is known
as the {\em error threshold}.

We shall now embark on these phenomena, starting with
error thresholds (they build on mutation and selection only);
we shall proceed with Muller's ratchet, which requires
 drift as well. Finally, mutational meltdowns
will be considered, which additionally require
an explicit model of population dynamics.
In contrast to common usage, we are reluctant
 to speak of `models of error thresholds' or `models of
Muller's ratchet'. Instead, we prefer to think about  models in terms of
their ingredients, and to examine which features they exhibit
in dependence of these ingredients.

As we shall see, clear definitions of the phenomena to be
examined are lacking
in places.
Things are clearest for mutational meltdowns but less so
for error thresholds. The phenomena may, however, be fairly well {\em
described}, if not defined, in terms of prototype models
which exhibit the respective behaviour. These are, not
surprisingly, the historical archetypes, and they all employ the
 permutation invariant class of fitness landscapes.

\subsection{Error thresholds}
They are the most prominent features of the
so-called {\em sequence space models}, as  inspired by the
discovery of the molecular structure of genes, and originally aimed
at prebiotic evolution in the RNA world.
The most well-known sequence space model
is Eigen's {\em quasispecies model} \cite{Eig71} which may
be understood as the coupled mutation-selection model
(\ref{coupled}), but with alleles replaced by sequences
$\bs{\sigma} \in \{+,-\}^L$, and
symmetric  mutation according to Eq.~(\ref{ss_mut}). That is,
the ODE system reads
\begin{equation}\label{quasispec}
 \dot  p_{\bs{\sigma}} = \sum_{\bs{\sigma '}} v_{\bs{\sigma \sigma'}}
    r_{\bs{\sigma'}} p_{\bs{\sigma'}} -
  \big ( \sum_{\bs{\sigma '}}  r_{\bs{\sigma'}} p_{\bs{\sigma'}} \big )
     p_{\bs{\sigma}}
\end{equation}
for the relative frequencies of sequences $\bs{\sigma}$.

\subsubsection{The prototype model}
Whereas sequence space mutation  is a natural mutation model for
the situation considered,
there is no canonical fitness landscape. The toy model which has
received a lot of attention is the SPL (sharply-peaked landscape)
with  selective advantage
$s$ of the wild type.
(The quasispecies community  has created its own terminology,
for instance, this favourable sequence is
called `master sequence'; we shall, however, avoid using
redundant terminology here.)

\begin{figure}
\hspace{0cm}
\begin{center}
\scalebox{1.}{\includegraphics{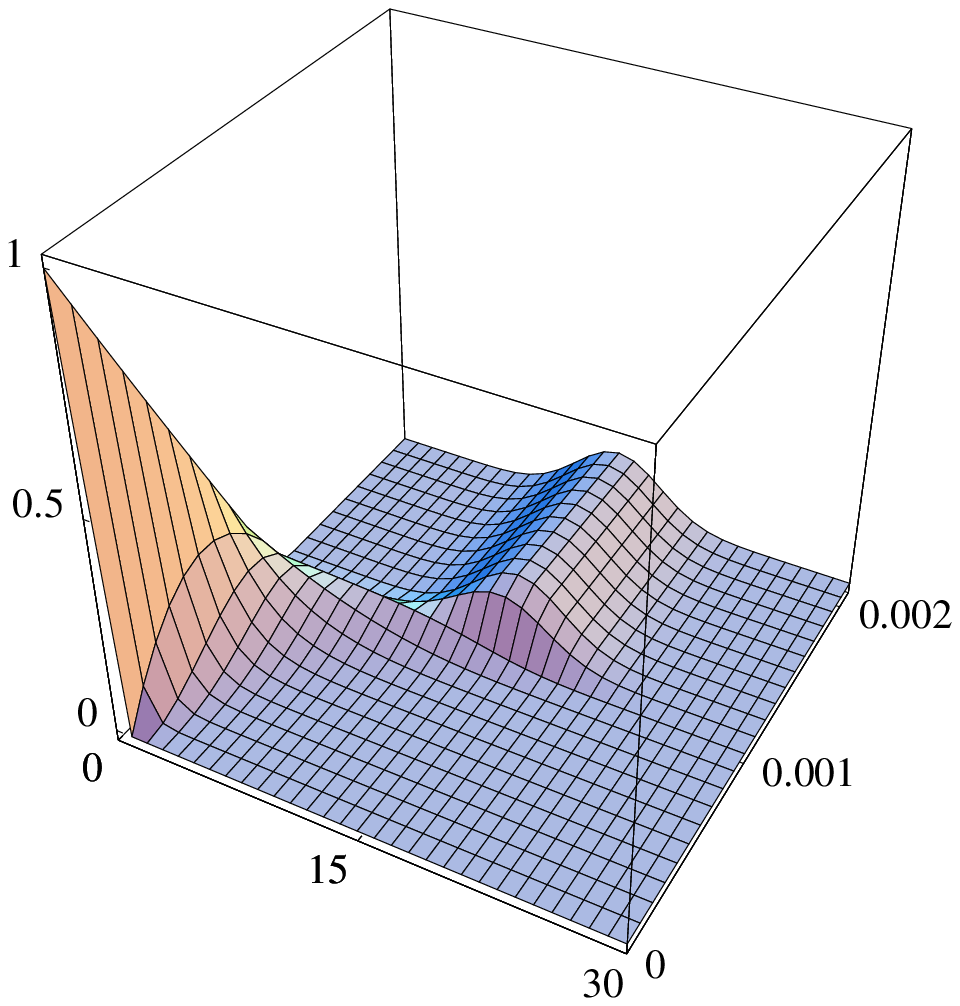}}
\end{center}
\caption{\label{spl}}
\end{figure}

This toy model
is so well-known because its stationary state
exhibits a behaviour reminiscent of a phase transition.
Whereas the population is closely centred around the favourable
sequence at small mutation rates, it is close to evenly distributed
over the space when $p$ surpasses a critical value. This
is shown in  Fig.~\ref{spl} for the closely related paramuse model,
to be described in the next subsection.
 This may be interpreted as mutation becoming
 so strong that it can no longer be counteracted
by selection, which, in due course, leads to the (effective) loss
of the favourable sequence. This  feature became
known as the {\em error threshold}.

The quasispecies literature up to 1989,
and the sharply-peaked landscape in particular,
has been comprehensively reviewed \cite{EMcCS89}, wherefore we
shall not dwell on details here. It is, however, important to note
that, in spite of the apparent simplicity of this
landscape, there is no exact  analytic solution known, not
even for the stationary state. One relies on numerical solutions,
either through integration of the ODE system, or determination of the
dominant eigenvector. Both take advantage from a reduced
representation of the mutation-reproduction matrix, which is available
thanks to the symmetries of the system. However, the special SPL situation
is readily approximated by neglecting back mutation
from the mutants to the favourable sequence, which are  very
rare events indeed. This yields the critical mutation rate \cite{EMcCS89}
\begin{equation}\label{critmut}
   p \simeq 1- \Big ( \frac{1}{1+s} \Big )^{\frac{1}{L}} \simeq \frac{s}{L}
\end{equation}
to first order in $s$, in good agreement
with the numerical value. It should be noted, however, that
a different approximation technique as proposed in \cite{AlFo96}, which treats
the sites as statistically independent entities, leads to
severe artifacts. For example, it predicts a maximum $s$
above which the error threshold
does not occur, no matter how large the mutation rate; such a feature
is absent from the original system.

Relation (\ref{critmut}) is readily interpreted as a
maximum mutation probability allowed at a given selective advantage
and sequence length, or, alternatively, in terms of a maximum sequence length
that may be correctly maintained under a given selective advantage
and mutation probability. Both may be taken as indicative of the
need for
the evolution of repair mechanisms in the course of the evolution
of the larger eukaryotic genomes \cite{MaySS95}. In invoking such arguments,
however, it is usually overlooked that the inverse relationship
(\ref{critmut}) relies
on the toy model of a sharply-peaked landscape as
generalized to whole genomes, which is, at best, dubious (see Sect.~4.2);
this point is discussed in detail in \cite{Wie97}.
It should be added that, although we
have not yet dealt with finite populations in sequence space,
 even very large populations
(viruses: $N \simeq 10^{12}$;  {\em Drosophila}:
 $N \simeq 10^{6}$) would be very
unlikely to find the isolated peak, even if
sequence lengths were tiny (say $L=1000$).

 On the other hand, the entire error
threshold phenomenon has occasionally been dismissed on the grounds
that it its based on an SPL, e.g.\ \cite{Charl90}, which is inadequate
as long as it is unknown which other landscapes exhibit this phenomenon, too.
Both lines of argument make clear that it is now imperative to study
other fitness landscapes.
Since there are only few clues as to what a biologically
relevant fitness landscape is, it is imperative to
study a variety of choices. But this is paved with obstacles
(recall that, even for the SPL, there is no exact analytical
solution). The only benevolent case is the Fujiyama (resp. multiplicative)
landscape (\ref{mult}),
where the independence of the sites allows for a straightforward
solution. The stationary state  has been given by several authors
\cite{Rum87,Higgs94,BBW97}. Namely, the frequency $p_j$ of sequences
with $j$ `$-$' sites follows a binomial distribution,
\begin{equation} \label{fuji_stat}
 p_j = \binom{L}{j} a^j (1-a)^{L-j}\,,
\end{equation}
where the parameter $a$ depends smoothly on the relative strength
of mutation and selection; in particular, $a=0$ for $p=0$, and
$a=1/2$ for $p=1/2$.
  Clearly, there is {\em no error threshold}; instead,
both the mean fitness and the genetic structure of the
population fade away gradually with increasing mutation rate.
For all other landscapes, more elaborate approaches are required.

We shall therefore proceed by discussing methods of analysis.
It will then be necessary to clarify more precisely what an
error threshold is supposed to be. After this excursion, we
shall summarize the results that have been obtained.
We shall finally discuss the experimental clues concerning the
error threshold phenomenon.

\subsubsection{Methods of analysis}
The approaches used are all from statistical physics; the
most important ones involve Ising models. Leuth\"ausser \cite{Leut86,Leut87a}
established an exact equivalence between a discrete-time
version of the quasispecies model, and a 2D classical Ising model.
To see this equivalence, consider  a 2D square lattice anisotropic Ising
system with Hamiltonian
\begin{equation}
 {\cal H}_{class} = \sum_{i=1}^{n} \Big ( E(\bs{\sigma}^i) -
\sum_{j=1}^L J \sigma_j^{i+1} \sigma_j^i \Big )
\end{equation}
with $\bs{\sigma}^i$ the spin configuration of the $i$'th row,
$E(\bs{\sigma}^i)$ its energy, and $J$ an interaction constant.
 The corresponding
 row-to-row transfer matrix (cf.\ \cite{Thom72})
 has elements
\begin{equation}\label{tm}
 {\cal T}_{\bs{\sigma}'\bs{\sigma}} =
 \exp \big ( - \beta E(\bs{\sigma}) \big )
 \exp \big ( \beta J \sum_j \sigma_j' \sigma_j \big )\,,
\end{equation}
where $\beta$ is the inverse temperature.
In close analogy, the elements of the mutation-reproduction matrix,
 ${\cal T}:={\cal VW}$
of Eq.~(\ref{absdiscretemuse}),
may  be written as
\begin{equation}
  {\cal T}_{\bs{\sigma}'\bs{\sigma}} =
  (p (1-p))^{L/2} \cdot \exp(r_{\bs{\sigma}})
  \exp \big ( \beta J \sum_j \sigma_j' \sigma_j \big )\,,
\end{equation}
where $\beta=-\ln \frac{p}{1-p}$ and $J=1/2$; note that
$(p (1-p))^{L/2}$ is a constant factor independent of the spin
configuration.

\begin{figure}
\hspace{0cm}
\begin{center}
\scalebox{.75}{\includegraphics{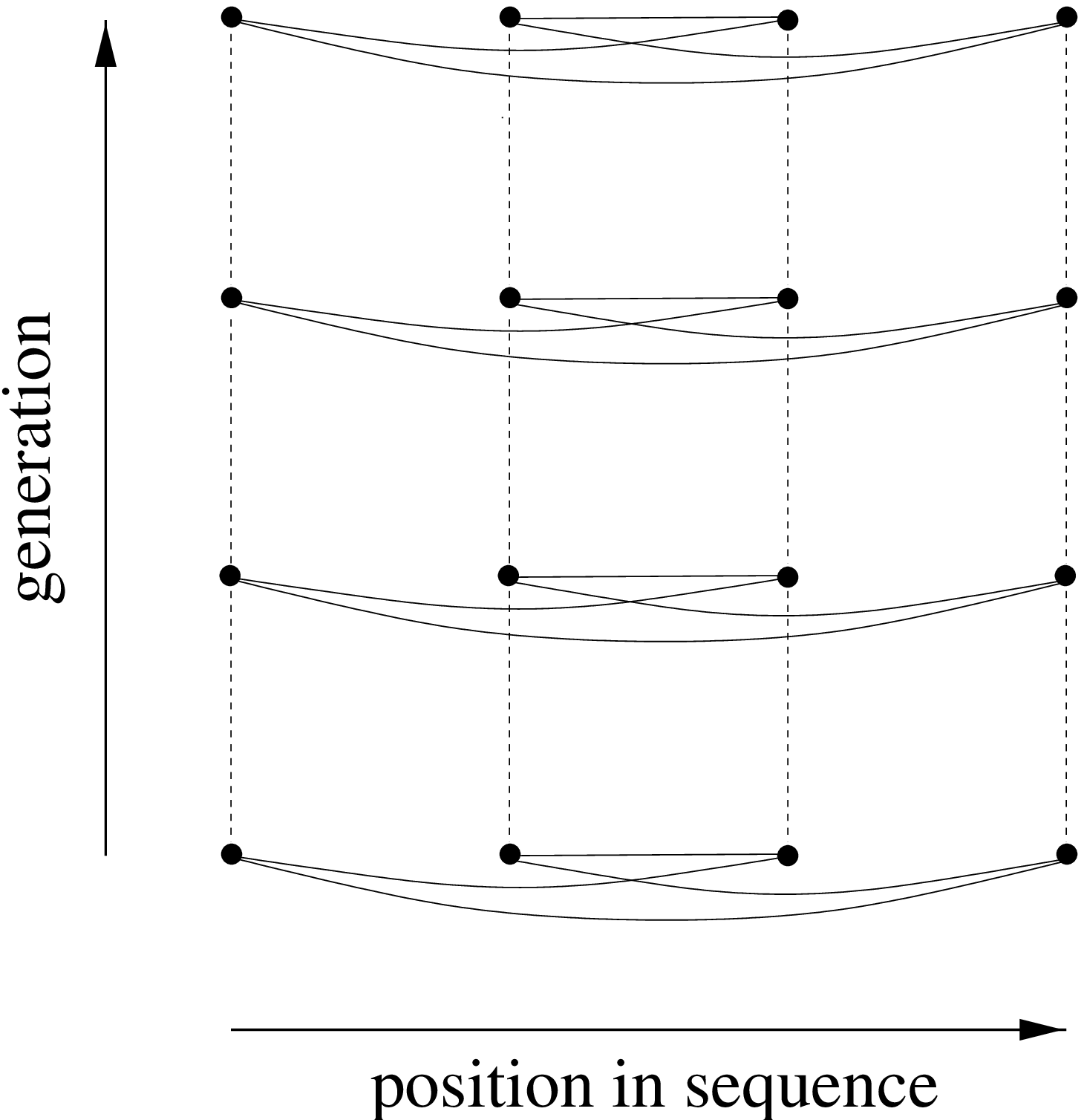}}
\end{center}
\caption{\label{ising}}
\end{figure}

The Ising system involved here is anisotropic: It has
nearest-neighbour interaction {\em between} the rows (which correspond
to mutation), but {\em arbitrary} interactions within the rows (the
within-row interaction energy corresponds to the fitness of the configuration,
the details reflecting the fitness landscape); see Fig.~\ref{ising}.

With this in mind, solving the evolution model is paramount
to diagonalizing
the transfer matrix of the corresponding Ising model. In particular,
knowledge of the
largest eigenvalue allows for the calculation of the
stationary state. This equivalence was exploited in a number of applications.
Tarazona \cite{Tara92} tackled the quadratic
(Mattis) landscape (\ref{quadr}), as
well as Hopfield Hamiltonians (\ref{hopf}); Franz and Peliti \cite{FrPe97} and
Franz, Peliti and Sellitto \cite{FPS93} examined the random energy model.
On the whole, however, results have been surprisingly sparse.
The reason seems to be that transfer matrices are hard to treat,
not just due to their size, but due to the characteristic
anisotropy of the interactions involved.

A related analogy which was described recently \cite{BBW97}
circumvents this problem.
It starts from the {\em paramuse model}, which had previously
been suggested as an alternative to the quasispecies model
\cite{Baa95}. It is the {\em para}llel {\em mu}tation-{\em se}lection
model (\ref{paramuse}) as adapted to sequence space,
\begin{equation}\label{paramuse}
 \dot  p_{\bs{\sigma}} =
    (r_{\bs{\sigma}} - \bar r) p_{\bs{\sigma}  } +
   \sum_{\bs{\sigma '}}  m_{\bs{\sigma} \bs{\sigma '}} p_{\bs{\sigma'}}\,.
\end{equation}
Here, the mutation rates simply read
\begin{equation}\label{paramut}
  m_{\bs{\sigma'} \bs{\sigma}} = \begin{cases}
\mu, &  d(\bs{\sigma'}, \bs{\sigma}   ) =1 \\
-L \mu, &    \bs{\sigma'} = \bs{\sigma}   \\
0, & \text{otherwise,}
\end{cases}
\end{equation}
where $\mu$ is the mutation rate per site.

With this, one arrives at
a mutation-reproduction matrix of the form
${\cal H} ={\cal M}+{\cal R}$ instead of
${\cal T}={\cal VW}$, which is exactly equivalent to the
 Hamiltonian of an {\em Ising
quantum chain}.
The limit which relates  Ising quantum chains to their classical
counterparts (cf.\ \cite{Kog79}) is just the short-generation limit which we have met
in Eq.~(\ref{shortgenlim}).

Explicitly, the quantum chain Hamiltonian reads (up to a constant term):
\begin{equation}
 {\cal H} = {\cal M} + {\cal R} =
 \mu \sum_{j=1}^{L} \sigma_j^x + \sum_{j=1}^{L} \rho_j \sigma_j^z
  + \sum_{j,k=1}^{L} \rho_{j,k} \sigma_j^z \sigma_k^z +
 \ldots \text{terms up to $L$'th order}\,,
\end{equation}
where, just for this moment, $\sigma^x$ and $\sigma^z$ denote Pauli's
matrices, and
\begin{equation}
\sigma^a_j :=  \bs{1} \otimes \ldots \otimes \bs{1} \otimes \sigma^a \otimes
  \bs{1}  \otimes \ldots \otimes \bs{1}
\end{equation}
an $L$-fold tensor product
  with $\sigma^a$  in the $j$'th place. Further, the collection of  $\rho$'s
determines the fitness landscape; in most cases, only terms up
to second order are involved (and  difficult enough to handle).

As a note of caution, let us remark that, although the Hamiltonian
is symmetric, the evolutionary dynamics does {\em not} have {\em
detailed balance}. This is because the linear ODE, $ \bs{\dot x} =
{\cal H} \bs{x}$, has no stationary state at all; the normalized
system (\ref{paramuse}) does have a stationary solution, but, due
to the nonlinearities involved, it does not fulfil the conditions
for detailed balance (i.e.\ $\partial
f_{\bs{\sigma}}(\bs{p})/\partial p_{\bs{\sigma'}} \neq
\partial f_{\bs{\sigma'}}(\bs{p})/\partial p_{\bs{\sigma}}$,
where the $f_{\bs{\sigma}}(\bs{p})$ constitute
the right-hand side of (\ref{paramuse})).

The toolbox developed for quantum chains
may then be applied to the evolution model. Some new
techniques are, however, required to take care of the
fact that it is not the quantum mechanical states which are
relevant for the system; the problem remains one of classical
probability \cite{BBW98}. So far, only two nontrivial cases have been worked
out in detail, namely Onsager's landscape and quadratic
fitness functions \cite{WBG98}.

Another recent approach is the mapping of the evolution problem
onto the Hamiltonian of directed polymers \cite{GGZ96,Gall97}.
Here, sequences are identified with elastic polymers, wandering in
sequence space directed along the time axis, and subject to
a potential. Here, mutation plays the role of elasticity,
and the potential determines the fitness landscape; this leads to
a transfer matrix very similar to (\ref{tm}).
So far, the only application seems to concern the SPL, which
corresponds to a {\em pinning potential} \cite{Gall97}.

With all methods of analysis which we have mentioned,
 analytical (exact or approximate)
studies are possible of order parameters and phase transitions,
but numerical simulations are often required to resolve the population
structure in detail. These, too, profit from the corresponding
methods in statistical mechanics, see, e.g., \cite{Tara92}.

\subsubsection{Characterization of error thresholds}

We have, up to now, carefully avoided to give a definition of the
notion of `error threshold'. Writing for a physical readership, we
have tacitly interchanged it against the concept of {\em phase
transition}, expecting the reader's approval. She or he may have
noticed, however, that this need not, a priori, go together with
the original descriptions.

Let us recall the original verbal description of mutational
degradation. Mutation
can no longer be counteracted by selection,  so that genetic information
is lost. This   implies both  a
genetic and a fitness aspect, and let us add that there may also be
an intermediate phenotypic aspect. But  there is,
as yet, no generally accepted definition of the error threshold
phenomenon.
Several descriptions are in use, but since error thresholds have
been so closely tied to the prototype model (SPL with intensive scaling,
and extensively-scaled symmetric mutation),  criteria have
been oriented towards this one, as well. The original criterion
was the loss
of the fittest sequence \cite{Eig71}; for the SPL (but not, necessarily, for other
landscapes), this
goes together with delocalization of the population over sequence
space. Although pictures like Fig.~\ref{spl} seem to speak a clear language,
 both properties are never met exactly with finite
$L$ for which the error threshold was originally described,
 but are expected to become exact only in the  limit
$L \rightarrow \infty$.
 In general, error thresholds appear to be strictly
definable only in this limit.
Mathematically speaking, this is because,
for finite $L$  and a primitive mutation matrix,
there is a stationary distribution with $p_{\bs{\sigma}} > 0$ for
all $\bs{\sigma}$;
hence nothing can be lost, and
no delocalization may occur. Physically speaking,
no phase transition is possible with a finite number of sites.
Note that the infinite population limit (\ref{infpoplim}) is implied
 in the description
by differential equations (\ref{quasispec}), and corresponds to
the thermodynamic limit of statistical physics.

Let us therefore summarize possible error threshold criteria, with
infinite sequences in mind. If the  landscape has a
single peak at sequence $\bs{\xi}$ say,
the loss of the fittest sequence, and the vanishing  of the
average overlap 
($u := \sum_{\bs{\sigma}} p_{\bs{\sigma}} \sum_i  \xi_i \sigma_i$) with it,
are the favoured criteria;
it should be noted, however, that they need not give identical
results \cite{FrPe97,Wie97}. With this kind of landscape,
$\bs{\xi}= ++++++\ldots+$ may be chosen  without
loss of generality; then, $u$ corresponds to the magnetization
of a classical spin system. To avoid confusion with the
corresponding quantum mechanical
quantity, we have previously termed it {\em surplus} \cite{BBW97}.
If there are multiple (but still isolated) peaks, as in the
 Hopfield landscape (\ref{hopf})
(they may stem from multiple patterns $\bs{\xi}^p$, as well as
  spin reversal symmetry), this order parameter performs
one or several bifurcations, each indicating the `loss of discrimination'
between a pair of peaks \cite{Tara92}.
If peaks are not unique, as with quantitative
traits or  RNA structures,
the populations may  spread over their neutral space,
without being delocalized in phenotype space. Error thresholds
are then reasonably defined as loss of the fittest phenotype;
 this is known as  {\em phenotypic
error threshold} \cite{HSF96,RFS98}.

The delocalization criterion is appealing, but problematic.
This is apparent from
the stationary state (\ref{fuji_stat}) of the multiplicative landscape.
Here, for every $j$, $p_j \rightarrow 0$ for $L \rightarrow \infty$
for any fixed but nonvanishing mutation rate.
 Hence, the distribution of genotypes
is {\em de}localized over sequence space however tiny the mutation
rate.  In particular,
the fittest sequence is lost. Nevertheless, one would not
 want to speak of an error threshold here (see above).

Another quibble with criteria based on knowledge of the fittest
sequence is that they, in a sense, represent
the standpoint of an omniscient observer. Due to the mixing
of a genetic and a fitness aspect, such quantities are not
observable independently of the  fitness
landscape. In contrast, the mean fitness of the population
{\em is} observable, at least in principle, as is its genetic structure.

In line with our previous wording, we therefore propose phase
transitions, or, alternatively, bifurcations of equilibria,
as criteria for error thresholds.
Actually, a general connection between
 bifurcations of equilibria and phase transitions (in the sense of
nonanalytic points
of the free energy) has been conjectured, although this is far from 
being proven
rigorously \cite[Ch.~5.7]{Ruel83}. Both phase transitions
and bifurcations go together with all threshold phenomena
described so far, and those to be described in the sequel.
On the other hand, they do not apply to the multiplicative
 landscape, as desired.
One observation concerning the SPL may be interesting in this context.
Although  the above considerations
are strictly applicable in the thermodynamic limit only,
the situation here may be mimicked
by a simple two-type model, where all unfit sequences are lumped
together and mutation is {\em unidirectional} (no mutation back to the
rare -- fit -- sequence) \cite{BaWi97}. In this model, a (transcritical)
bifurcation occurs with the mutation rate as the bifurcation
parameter, in the course of which the fittest sequence
is lost.  
%It is instructive
%to observe how  this bifurcation arises from an
%`avoided crossing' of the two equilibria in a family of models
%where the mutation changes from symmetric to unidirectional;
% see Fig.~\ref{avcro}.

\subsubsection{Results}

Phase transitions in the usual sense
may only occur if fitness and mutation both scale extensively,
or both scale intensively, cf.\ \cite{FrPe97}. Then, the result
is a critical mutation rate per site, or per genome, respectively.
 If, on the
other hand, fitness scales intensively and mutation extensively,
one may obtain an inverse relationship between sequence length and
mutation rate, as, for example, the one given in (\ref{critmut}).

In the permutation-invariant quadratic landscape (\ref{quadr}),
a phase transition (of second order) is present  if $\alpha=0$,
$\gamma < 0$ \cite{BBW97}. In this case, fitness is a convex
function of the surplus, or, put differently,  epistasis is negative.
In contrast, lack of epistasis ($\gamma=0$)
precludes phase transitions; see our discussion of the Fujiyama
(resp.\ multiplicative) landscape in the previous section.
The same is true for positive epistasis
($\gamma>0$) \cite{Charl90,Wie97}.
These observations agree well
with classical results predicting a higher mutation load
in situations with negative epistasis as compared with positive
epistasis \cite{KiMa66}.
 With Onsager's landscape,
one observes a second-order phase transition, too. In contrast to the
quadratic landscape, where both surplus and mean fitness vanish
at the critical mutation rate,
the mean fitness continues to decrease
beyond the nonanalyticity point in Onsager's landscape \cite{BBW97}.

In Hopfield's landscape (\ref{hopf}), only the vicinity of the highest peak
is populated at small mutation rates.
With increasing mutation rate, secondary maxima (which are less high
but more abundant) take over, before, finally, the genetic
structure is entirely lost. In line with this, there is not a
single error threshold, but a sequence
of bifurcations,
each of which indicates the loss of discrimination
between two patterns of the neural network \cite{Tara92}.
Presumably, such a behaviour is typical of multi-peaked
landscapes.
 The importance of such entropic
effects can be mimicked by a toy model with one high and narrow,
and a second less tall but broader fitness peak \cite{SS88},
the bifurcation structure of which may be analyzed exactly
\cite{BaWi97}.

At the rugged end of the landscape zoo,
 the random energy model
was investigated  \cite{FPS93,FrPe97}. Locally, i.e.
in the vicinity of the highest peak, it was found to behave like
an SPL; this is attributed  to the complete lack of
correlation.

Both Onsager's and Hopfield's landscapes may be considered as
displaying (some suitable generalization of)
negative epistasis. This is due to the fact that on average,
with increasing distance from the maximum, a larger fraction of
additional mutations either do no further harm, or even act in a compensatory
manner. One might be tempted to conjecture that this type of epistasis is
required for a phase transition to occur. It would be
interesting to know the behaviour of the MQT (multiple quantitative
traits) model (\ref{mqt}), which
may be interpreted as displaying something like
{\em positive} epistasis (in the same
way as quadratic landscapes with stabilizing selection do),
but this has not yet been examined.

\subsubsection{Error thresholds in finite populations}
Unlike with infinite populations, we do not even attempt to define
error thresholds for finite populations. Error thresholds
are strictly definable only in the infinite population limit.
 However, obvious dramatic
changes take place with finite populations as well, and we
shall restrict ourselves here to describing instead of defining them.

In turning to finite populations, we leave the realm of difference or
ordinary
differential equations. Given a fitness landscape and a mutation
model, the deterministic dynamics must be replaced by Wright-Fisher
sampling according to Eq.~(\ref{wf_sampling}), or a
corresponding master equation. Before presenting the results,
we go on a little methodological excursion which will also apply
to the later sections on finite populations.

\bigskip

{\bf Methods of analysis:}
For all but very simple selection schemes, it is the hour of the
Monte Carlo'ist, especially if dynamical aspects are also considered.
Interestingly, although the stochastic equations are more difficult
analytically than their deterministic counterparts, they are
simpler to handle in simulations. This is because, unless the
symmetries of the fitness landscape can be used to advantage
(as in, e.g., \cite{Tara92,Peck94}), an exceedingly large number of
configurations must be dealt with in the deterministic case,
whereas the number of states to keep track of at any instant is
limited by the number of individuals in a finite population,
which is usually much smaller.

If  generations are discrete and the stochastic
component is introduced by Wright-Fisher sampling
(\ref{wf_sampling}),
the corresponding simulations are straightforward.
For large populations, the multinomial sampling is quite time
consuming, but may be sped up in various ways, e.g.\ by the
sampling scheme in \cite{GaBu99}
or the multinomial algorithm in \cite[p.559]{Devr86}.
 However, many
applications instead use the corresponding continuous-time
formulation via a master equation.
It lends itself directly to Monte-Carlo simulations.
Many authors explicitly reference the paper by Gillespie
\cite{Gill76}, which gives a very nice and ready-to-use exposition to
simulation methods taylored for stochastic chemical reaction systems
(note that the population genetic equations may, indeed, be formally
understood as reaction systems in a flow reactor; this is actually
the context Eigen rediscovered them in \cite{Eig71}).

For simple selection schemes, some analytical approximations
 of the stochastic equations are possible. They often
employ moment expansions \cite{Bue91} for various quantities
of interest. To be
more precise, both moment \cite{HiWo95} and cumulant \cite{PBS97}
expansions have proved useful. The notorious problem with
both methods is the fact that lower-order moments or cumulants
depend on the respective higher-order quantities, and some kind
of closure approximation must be made. Possible choices include
the use of higher moments derived from the corresponding deterministic
equations \cite{HiWo95}, as well as maximum entropy considerations.
The latter
approximation is not very satisfactory, however, if mutation,
selection and drift are the only evolutionary forces considered
\cite{PrBe97}. It works much better when recombination is present,
too (recombination is a main ingredient of genetic algorithms, for
which the method was originally developed; for review,
 see \cite{PBS97}). This is
because recombination reduces the higher-order cumulants and,
with them, the sensitivity of the solution towards them \cite{PrBe97}.

\bigskip

\begin{figure}
\hspace{0cm}
\begin{center}
\scalebox{1.}{\includegraphics{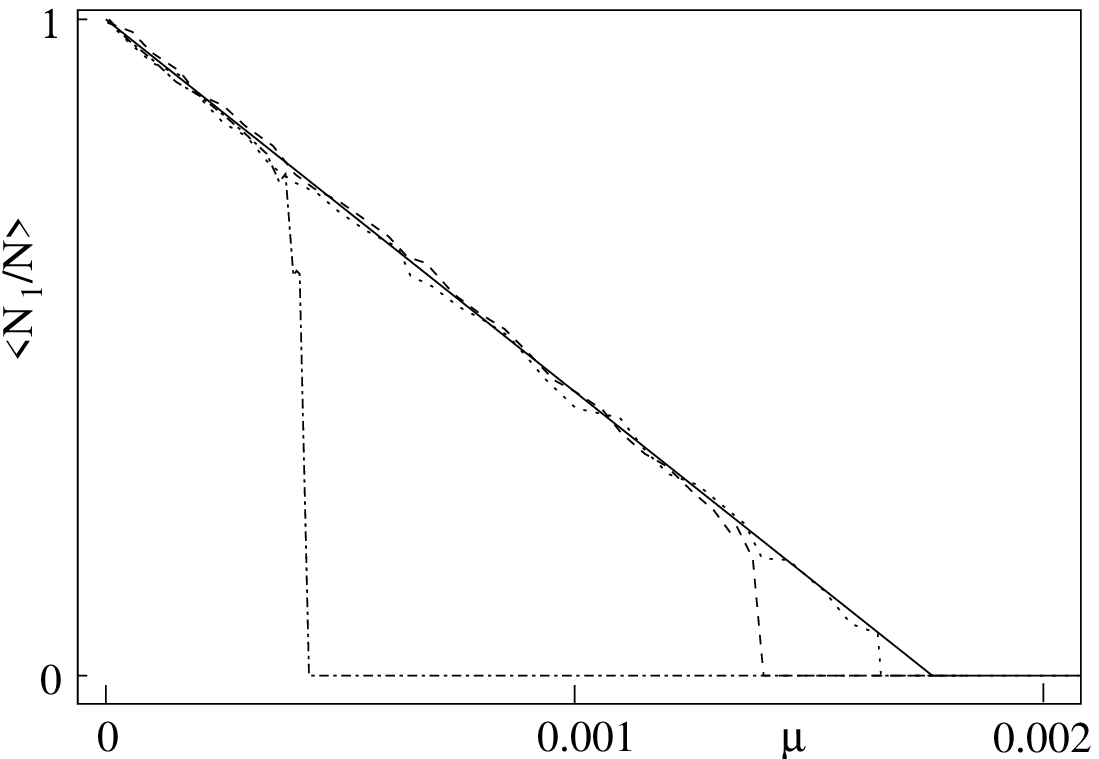}}
\end{center}
\vspace{5cm}
\caption{\label{stochet}}
\end{figure}

{\bf Results:} Error thresholds in finite populations were first described by
Nowak and Schuster \cite{NoS89} for the sharply-peaked landscape.
Here, the fraction of advantageous sequences is the relevant
random variable, and its expectation and variance are of primary
interest. In simulations, they may be measured as long-time
averages.
Interestingly, the expectation
follows the deterministic curve
for small mutation rates, but then `jumps' to (near) zero;
 see Fig. \ref{stochet}. This could also be corroborated
analytically \cite{WBS95}. The
transition is characterized by large fluctuations. Beyond this point,
the population is indistinguishable from a finite population
on a flat landscape.

More precisely, finite population size shifts error thresholds
to lower mutation rates by an amount which seems to be roughly proportional
to $1/\sqrt{N}$ \cite{NoS89}. Apparently,  the deterministic solution may
be interpreted as the time average of the stochastic process
for mutation rates outside
 the `window'
between the deterministic and stochastic transitions,
but this is not so within the window.

Very similar observations hold for the corresponding  SPL
in the RNA world \cite{RFS98}, where one structure is distinguished
over all others. For small mutation rates, the
favourable phenotype
is conserved, while the genotypes diffuse through the corresponding
neutral network. At some critical mutation rate, a {\em phenotypic
error threshold} takes place \cite{HSF96,RFS98}.

Simulations were also performed for the Sherrington-Kirkpatrick
 spin glass \cite{BoSt93}.
In line with the deterministic results for the closely related
Hopfield's landscape,
three regimes
emerge here. For small mutation rates, the population remains
stationary in the vicinity of one high peak; for intermediate
mutation rates, it starts wandering across secondary (but
more abundant) peaks; and for high mutation rates, it diffuses
through all of sequence space in the long run.

\subsubsection{Clues from the real world}
We have, so far, addressed error thresholds as `phenomena',
thereby implying they are real. However, we have not yet faced
the question whether they are relevant problems
for evolution.
Today's species clearly exist and are
genetically well-defined entities.
The question must therefore be rephrased to read: Has the
mutation rate of these organisms evolved to avoid the error threshold,
or didn't it have to bother?
It is widely believed that the former is the case, but the
experimental evidence  is not yet conclusive.
 Let us follow the  clues.

An indirect piece of evidence comes from comparison of
genome sizes and mutation rates across species.
A roughly inverse relationship is observed for quite a
variety of organisms and genome sizes, see the recent
survey by Drake at al \cite{DCCC98}.  This was
taken to indicate that the mutation rate has evolved
to avoid the error threshold \cite{JMS89} as given by
Eq.~(\ref{critmut}). However, this conclusion implicitly
relies on the assumption that the SPL is the relevant 
landscape for the whole  genome; the reservations concerning
this assumption have been considered in Sect.~4.2.

More direct evidence comes from
 mutagenesis experiments with RNA viruses. These viruses
have very large genetic variability even at their natural
(spontaneous) mutation rate, as reviewed in \cite{DoHo88}.
If their mutation rate is increased with the help of chemical
mutagens, the fitness of the population decreases. The
virus does not survive mutation rates larger than twice or
three times the spontaneous one, presumably
because the error threshold is surpassed; see \cite{HDTS90}
 and \cite{DESM+96,DoHo97}
for reviews.

Of course, such measurements
are rather crude and do not give any hints at the details of the
phase transition. More detailed information can only be gained
from observation, and possibly sequencing,
 of large samples, taken from large populations
 under stationary conditions and a variety of mutation rates.
With the power of sequencing methods increasing almost daily,
this might not be out of reach for viruses
 or populations of
{RNA} molecules which may be replicated in the lab under
tunable mutation rates; see \cite{BiGa97} for a review.
If such observations become available, they will tell us
 a lot about fitness landscapes.
After all, we have seen that some fitness landscapes have error
thresholds, whereas others do not.

\subsection{Muller's ratchet}
The prototype model which displays Muller's ratchet is
aimed at the multilocus context in {\em finite} populations.
It describes the fate of a finite population which is released at the
peak of a multiplicative fitness landscape and experiences
deleterious mutations, as well as genetic drift. To be more precise,
an infinite number of sites is assumed, each of which may be
`wildtype' ($+$) or `mutant' ($-$), and mutation occurs in a unidirectional
fashion (from $+$ to $-$) at genomic mutation rate $U$, i.e.\
the infinite sites limit is assumed.
The population is haploid (or diploid without dominance), and the
fitness of an individual with $j$ mutations is $w_j=(1-s)^j$. A population
of size $N$ undergoes Wright-Fisher sampling (\ref{wf_sampling}) at
every discrete generation.

Since we have a permutation-invariant landscape,
it is sufficient for most considerations to
consider classes of individuals with the same number of mutations.
Under the assumptions made, the probability of the $j$'th
mutation class to be sampled reads
\begin{equation}\label{psi}
\psi_j = \frac{1}{N} \sum_{k=0}^{j} v_{j,j-k} \frac{w_{j-k}}{\bar w}
n_{j-k}\, \quad
\text{where} \;  v_{j,j-k}=\frac{U^k}{k!} e^{-U}\,;
\end{equation}
cf.\ Eq.~(\ref{wf_probs}). Alternatively, it may be assumed that
sampling takes place right after selection (instead of after
mutation), in which case one has
\begin{equation}\label{psi_alt}
\psi_j = \frac{1}{N} \sum_{k=0}^{j} \frac{w_j}{\bar w} n_j v_{j,j-k}\,.
\end{equation}

The qualitative behaviour of the model is intuitively clear.
Under the action of mutation, some individuals will
soon acquire mutant sites, even if the whole population was
 initially free of mutations.
If then, due to the hasards of the sampling process, no individual with zero
mutations becomes mother in the next generation,  the zero
mutation class is lost from the population. This process repeats itself
because the then actual fittest class will have the same fate as
the zero mutation class. The mean fitness of a clonal lineage will
decline incessantly by the successive loss of the actual
 least loaded classes. Due to the unidirectional nature of
mutation, `better' genotypes can never be reestablished;
thus, the mechanism is irreversible and proceeds in a ratchet-like
manner.

 Recall from Sect.~4.3 that, in the infinite sites limit, novel
mutations will always occur  at different sites. For large
but finite $L$, this is still the case for most mutations.
Accumulation of deleterious mutations may thus occur without
necessarily invoking fixation of specific configurations themselves
\cite{HiWo95,ChCh97}. Here, a fundamental difference between
sexual and asexual reproduction becomes apparent. If
recombination were present, the fittest genotype could be re-established,
except in the rare case that a specific site is fixed
 for a  mutation. Therefore, the ratchet-like deterioration process
is much more pronounced with asexual than with sexual reproduction.
This was first pointed out by Muller (1964) \cite{Mul64}, and
Felsenstein (1974) \cite{Fel74} named the process Muller's ratchet.

\subsubsection{Ratchet dynamics}

It is instructive to consider, for a comparison, the corresponding model
of mutation and selection in an {\em in}finite population,
with {\em finite} $L$, and  {\em symmetric}
mutation according to Eq.(\ref{ss_mut}).
We have met this before in the context of sequence
space models, and the stationary state (i.e.\ the distribution of genomes
with $j$ mutations) was given by Eq.~(\ref{fuji_stat}).
In the infinite genome limit (\ref{infgenlim}),
this converges to a Poisson
distribution with parameter $U/s$ (this holds
irrespective of whether mutation at single sites is symmetric
or unidirectional, provided the population
was initially released at the fitness peak).
 This is termed the {\em deterministic
limit} in the  literature on Muller's ratchet, and much of the theory is
based upon it,
cf.\ Haigh (1978) \cite{Haig78}. If, instead of (\ref{psi}),
(\ref{psi_alt}) is used,  one obtains a
Poisson distribution with parameter $U (1-s) /s$, instead of $U/s$
\cite{GLB93}. This may be more realistic in certain cases, and
we shall adhere to it in what follows.

Let us now move on to finite population size, which will destroy
the stationarity of the solution. The crucial events in the process
are the losses of the actual least loaded classes -- these are
known as `turns' of the ratchet. It is, therefore, important to
have good estimates of the rate of the ratchet as a function of the
parameters $N$, $s$, and $U$. A crucial quantity is the relative
size of the least loaded class. In the deterministic limit, this is
\begin{equation}
p_0 =  \exp \Big ( -\frac{U (1-s)}{s}
                       \Big )\;.
\label{eq:c0gabriel}
\end{equation}

Equation (\ref{eq:c0gabriel}) may be used to obtain a very rough
estimate for the rate of ratchet: in every generation, it turns with the
probability $P$ that all $N$ sampled mothers  of the next generation
do not belong to the mutation-free class, therefore,

\begin{equation}
P = ( 1 - p_0)^N.
\label{eq:Pratchet}
\end{equation}

This estimate is accurate only if the process is very slow.
For, in this case, the  distribution of the
mutation classes is
restored to the deterministic expectation between any two turns.
 But even when it is not accurate, it gives some qualitative feel
for the ratchet dynamics.
Large population size decreases the rate of the
ratchet and does so efficiently if $U (1-s)/s$ is not
too large. The smaller  $s$ or the larger $U$, the faster turns the ratchet.

If the process is very fast (that is, if $N p_0<1$, perhaps  because of
high mutation rates),  the ratchet can be treated
as a quasi-deterministic process (for details see \cite{Gess95}).

Many attempts have been made to obtain good estimates of the rate of
the ratchet
\cite{Haig78,PNL87,Bell88, GLB93, LBBG93,SCS93,Butch95,Gess95,HiWo95,ChCh97},
but all approximations are valid only for
restricted parameter regimes. The problem is astonishingly
difficult for one so simply defined.  The
main problem is that, under conditions where the ratchet
rate is reasonably large, the shape of the expected distribution
deviates considerably from the deterministic limit and is hard
to predict in a generally valid form. In particular,
no simple scaling between $U$, $s$, and $N$ seems to exist \cite{HiWo95}.

\subsubsection{Error thresholds versus Muller's ratchet}
Error thresholds and Muller's ratchet  have a lot in common:
They both describe mutational degradation;  in the prototype
models, this involves delocalization and, in particular, loss of the
fittest genotype. Both effects require infinite $L$. However,
error thresholds may be present
in infinite populations, whereas Muller's ratchet requires
the stochastic component brought about by finite population size.
After all, delocalization occurs in the SPL beyond a critical (small)
mutation rate, whereas the Poisson distribution
which emerges in the deterministic limit of the Muller's ratchet
model is
localized for {\em any} $U < 1$.
What is the crucial difference?

Wagner and Krall \cite{WaKr93}
approached the question on the basis of a general model class
 with infinite population size,
discrete generations, unidirectional mutation, and the infinite
genome limit (\ref{infgenlim}).
Fitness landscapes come from the permutation invariant class,
with fitness $w_j$ decreasing monotonically with $j$,
the number of `$-$' sites. The population initially consists
of wild type individuals only. In line with the
classical localization results (\cite{Mor76,King77}; for review,
see \cite{Bue98}) the authors show
that the following are equivalent:

a.) The sequence of fitnesses $\{w_j\}$ has no positive lower
limit.

b.) For $U<1$, there is a
stationary distribution, with a nonvanishing frequency of
the fittest genotype.

Let us remark that a.) requires extensive scaling of fitness.

The result is plausible in that it stresses the purging effect
of strongly deleterious mutations. Put differently, a.) entails
that a few alleles are sufficiently advantageous to prevent
the population from spreading `too thinly' over the entire
genotype space.

The SPL, as the prototype landscape for error thresholds, does not
fulfil a.), whereas the multiplicative fitness scheme of
the Muller's ratchet model
 does.
As a consequence, the fittest genotype is lost for some $U_c < 1$ with
 the SPL, but
not for the multiplicative scheme, as long as the infinite genome limit
is assumed.  This explains the need for stochastic effects in the
prototype model for Muller's ratchet.

As we have discussed above, however, some care must be exercised
 in transferring results
relying on the infinite genome limit
to  sequence space models.
After all, the latter rely on the {\em infinite population limit}
(\ref{infpoplim}), with its {\em extensive} scaling of mutation.
In order to illustrate
 this point, let us reconsider the multiplicative
fitness function. We have seen that the population remains {\em
localized} in the {\em infinite genome} limit, as also predicted by
a.) and b.). On the other hand, it is {\em delocalized} in the
infinite {\em population} limit, cf.\ Eq.~(\ref{fuji_stat}) and the
discussion in section (5.1.3). Thus, the
different limits implied, together with the scaling assumed, are
also important for differences between error thresholds and
Muller's ratchet. However, no general characterization is available
at this stage.

\subsubsection{Mutational melt-down and extinction}

So far, we have assumed that population size remains constant,
even throughout the course of fitness deterioration.
Constant population size irrespective of the mean fitness
 is built into the standard
sampling process (\ref{wf_sampling}), which is invariant under
multiplication of all (Wrightian) fitness values with a constant. As long
as the number of potentially viable offspring exceeds the
carrying capacity of the biotope, this is a reasonable assumption.
At some stage of the ratchet process, however, this will no
longer hold. Then, {\em absolute} fitness values will become
relevant, and  population size will start to decline.

Models of this process were first considered by
Lynch and Gabriel (1990) \cite{LyGa90}.
They require an explicit model of population dynamics in addition
to mutation, selection, and drift. In essence, one assumes that
the expected number of offspring per individual, $R$, is independent of
 fitness and so large
that the number of offspring exceeds the
carrying capacity $C$ of the biotope at the beginning of every
generation. Then, offspring are allowed to survive probability $w_j$,
$0 \leq w_j \leq 1$ (viability fitness). Finally,
population regulation (e.g.\ according to a modified logistic equation)
brings the population back to a size less than or equal to $C$.

Of course, every population will finally die out  due to chance
effects, even without deleterious mutations, cf.\ \cite{GaBu92,Pal96}.
However, the long time scales involved will not be our concern here.
With the assumptions on selection, mutation and drift
as in the Muller's ratchet prototype model, mean fitness 
declines at constant rate until the expected number of viable
offspring falls below the carrying capacity (or, equivalently,
 the expected number of
viable offspring per individual, $\langle R \cdot \bar w \rangle$,
falls below 1). Then the population goes to extinction rapidly
because a gradual reduction of population size accelerates the rate
of the ratchet. This process is known as {\em mutational meltdown}.
Large fluctuations of population size occur in the vicinity of the
extinction point; the extinction times themselves, however, show
astonishingly little variation for given sets of parameters
\cite{LyGa90}. As to parameter dependence, extinction times depend
on the time course of fitness deterioration, and this, in turn,
depends on the rate of the ratchet as well as on the damage per
turn. A change of the detrimental mutational effect $s$ has
opposite effects on these. On the one hand, an
 increase in $s$ implies that selection is more efficient in removing
 new mutations from the population and the rate of the
ratchet is reduced; on the other hand, the damage per turn
 increases.
At small values of $s$, the second effect predominates
because the reduced rate of the ratchet is overcompensated by the
fitness loss per turn  \cite{LyGa90}. An intermediate mutational
effect $s_{\text{min}}$ minimizes the time
to extinction (or maximizes the extinction risk).
For $s > s_{\text{min}}$, selection becomes so efficient in removing
deleterious mutations that the extinction risk declines \cite{GLB93}
(after all, if all mutations are lethal, the ratchet will not turn at
all). The dependence of the extinction risk on $s$ is quite strong:
At $s_{min}$, the
time to extinction is reduced by several
orders of magnitude. 

Although we set out to exclude recombination from our considerations,
we should like to add that mutational meltdown  also endangers small
sexual populations (up to a population size of about 1000)
\cite{GBL91,Lan95,LCB95b,Bern96}. A contrasting result,
predicting recombining populations larger than 100 to be safe from 
Muller's ratchet \cite{CMC93}, is attributed \cite{LCB95a} to the lack
of explicit population dynamics in the simulations.

%A similar effect occurs also under sexual
%reproduction but only at very small population sizes where the
%fixation probability of deleterious mutations is high; this is
%discussed in \cite{GBL91,LCB95b,LCB95a,Lan95}).

\subsubsection{How to escape Muller's ratchet}
In the prototype model, Muller's ratchet, and mutational meltdown
as its consequence, are unavoidable. But are they generic features
of models of mutation, selection and drift?

In the prototype model, the ratchet  may be slowed
down  by low
mutation rate, large population size,  and large values of $s$.
But there are less obvious possibilities to retard the ratchet
when the assumptions of the prototype model are slightly relaxed.
If, instead of equal mutational effects at all sites,
a distribution of effects is assumed,
longevity is enhanced in a very
pronounced way, although the ratchet is not entirely halted \cite{LyGa90}.
The reason is that, even though the effects of mutation
are deleterious on average, the variance of the mutational effects introduces
the possibility that the reduction in fitness caused by a mutation
at one locus may be compensated by a beneficial
mutation at another locus.

Along another line of thought, positive epistasis  was assumed
and shown to halt the ratchet entirely \cite{Kond94}. However,
this  seems to be a nongeneric situation since it is reverted
when a distribution of (unconditionally deleterious)
effects is additionally introduced \cite{Butch95}.

But there are other possibilities to escape the ratchet once one relaxes
the assumptions of the prototype model. Instead of the multiplicative
fitness function with unidirectional mutation,
Wagner and Gabriel \cite{GaWa88,WaGa90} considered a model
of quantitative genetics with multivariate Gaussian mutation
and selection, as introduced in Section 4.2. In this landscape, the  proportion
of compensatory mutations increases with the distance
from the peak. Consequently,  the population will, during the
process of sliding off the peak,
finally arrive at a point where compensatory mutations become
 relevant, and
eventually halt the sliding.
A similar result holds for the multiplicative fitness function
when mutation is changed from unidirectional to symmetric \cite{HiWo95}.
Recall that unidirectional mutation is a perfectly
reasonable assumption in the infinite sites limit, even if
mutation is originally symmetric, provided the population is
 at the fitness peak.
If, however, back mutations are at all present in the original
model, they will eventually become important at some distance from
the peak.

In both situations, the population reaches a
mutation-selection-drift equilibrium, at which it moves in a
spherical shell centered at the fittest genotype (or phenotype)
\cite{WaGa90,HiWo95}. The radius of the shell increases with 
the mutation rate, but remains stable on a time average. Thus, whereas the
decrease of fitness ceases (and Muller's ratchet is halted), the
turnover of genotypes continues. If the population is released
beneath its equilibrium fitness, it will even climb towards that
equilibrium,
 a process favoured by higher mutation rates --
in such cases, a substantial proportion of the occurring mutations
is  beneficial.

We  are thus back at a question about fitness landscapes: How
abundant are compensatory mutations?

\subsubsection{Clues from the real world}
Clonal lineages arise frequently in both the animal and the plant
kingdom. Their life spans are fairly short (on the order of
$10^4-10^5$ generations) and  show surprisingly little variation
(see \cite{LyGa90} and references therein), in line with the
 predictions of
the prototype model. On the other hand, there are a few obligate
clonal species which are very old. One example are bdelloid rotifers,
tiny multicellular organisms which endure harsh environmental
conditions \cite{PoRi92}. If Muller's ratchet were the general explanation for
the extinction of clonal lineages, why could these species be an
exception? One might speculate that, due to the selective pressures
acting on them, $s$ is very large. But other `halting' mechanisms
might be considered as well.

The stochastic process of fitness loss over generations was be directly
observed in so-called serial passage experiments with viruses,
where  samples are repeatedly transferred from culture
to culture \cite{Chao90}. In these experiments, 
population size was reduced to the extreme case of
one individual between passages. One should be reluctant to
speak of   Muller's ratchet in this extreme situation -- it
simply demonstrates that deleterious mutations do occur (Brian
Charlesworth, personal communication). 
Nevertheless, fitness loss is also observed with less severe
population bottlenecks, and the dependence on population
size has been examined; this is
reviewed by Domingo and Holland \cite{DoHo97}.

\subsubsection{Error thresholds and mutational meltdowns}
We have, so far, considered explicit population dynamics, and
mutational meltdowns as its consequence, in the context
of Muller's ratchet only. It is a logical next step to examine them
for the prototype model of error thresholds, too. Malarz and Tiggemann
\cite{MaTi98} introduced explicit
population dynamics into a finite population on an SPL, together with
sequence space mutation (note that this includes back mutation to
the fittest type).

In line with what we know from  error thresholds in finite
populations (Sect. 5.1.5), there is a stationary distribution of types
 provided
the mutation rate is small. As a consequence, mutational meltdown, too,
can only occur when a critical mutation rate is surpassed.
Malarz and Tiggemann compared the parameter dependence of this
meltdown process with that of the corresponding stochastic error
threshold. Perhaps not surprisingly,
the  meltdown point will only agree
with the
stochastic error threshold in the unrealistic case
in which the expected number of offspring
matches the carrying capacity even in a population which has lost all
favourable sequences. Otherwise, meltdown takes place at
smaller mutation rates \cite{MaTi98}. In any case,
 one meets the typical meltdown curves
with accelerated meltdown rate and large fluctuations around the
extinction point. The equilibrium structure of the model was
corroborated by bifurcation analysis of a toy model in \cite{BaBe98},
see also Bagnoli and Bezzi, this volume.

\section{Connections with the molecular evolutionary process}

Much of what we have been concerned with so far was related to
equilibrium considerations and/or the time course of (mutational)
degeneration. However, some recent interest has turned to the dynamics
of adaptation, and to the time course of molecular evolution
(i.e. the turnover of genotypes) once a well-adapted state
has been reached. Much of the concepts and methods discussed so far also
lend themselves to the study of dynamical questions.

One recent study of adaptation dynamics concerns RNA structures.
Working with folding algorithms to assign the phenotype
and taking selective disadvantage to be proportional to the distance
from some target sequence, Huynen et al. \cite{HSF96} demonstrated
that evolution proceeds in an intermittent fashion  related
to the underlying landscape (i.e.\ neutral networks of frequent structures
percolating sequence space and penetrating each other). Long
 phases of constant phenotype, during which the population
diffuses in the underlying neutral space, alternate with
selection-induced, sudden transitions at positions where networks
come close to each other. These may be termed {\em adaptive sweeps}
in the language of population genetics, because they
temporarily eliminate  all molecular variation.

Intermittent behaviour, including hovering about metastable states,
 seems to be typical of the time evolution on
rugged fitness landscapes in general; see, e.g.\ the analysis of the
random energy model by Zhang \cite{Zhang97}. In general, if the
interval of observation is long enough, a biphasic behaviour
seems to be common. A population released at some random point
first enters a fast mode with large jumps in fitness; this is
hardly affected by population size. Later, it enters a noise-assisted
mode with small, rare jumps, for which the finite population
size is the driving force. Such or similar observations were
reported for the random energy model \cite{Zhang97},  for
the Sherrington-Kirkpatrick spin glass \cite{APS89}, and for the NK model
 \cite{Ohta98}. Here, we meet the old population
genetics rule of thumb (cf.\ Section \ref{ext}):
 For a new allele with selective advantage
$s$ over the  fitness of the existing population, selection is in effect
if $Ns>1$; otherwise, random effects predominate.

This `noise-assisted' phase is the crucial one for considerations of
molecular evolution in
the long run, with comparisons of different populations or even
species  in mind. In this
context, the loss of correlation between sequences in the course
of their divergence from a common ancestor is
a quantity of prime interest --  after all, this is the observable
when sequences from different species are compared.
For a strictly flat landscape as originally proposed for the
neutral theory, the behaviour is
well-known but is at variance with the observations, see \cite{Gill91}.
The first not-strictly-neutral
models considered were Ohta's {\em shift models}, assuming an
unchanging fraction of deleterious mutations to be available
over evolutionary time, with advantageous mutations
so rare that they do not contribute significantly to molecular
evolution; this is the {\em nearly neutral theory} as reviewed in
\cite{Ohta98}.
These models owe their name to the fact that
incessant fixation of deleterious mutations and, as
a consequence, an incessant decline in mean fitness (over phylogenetic time!)
is the invariable
consequence -- a Muller's ratchet like process!
For precisely these reasons, the shift models did not stand up
to scrutiny and were later replaced by the house-of-cards model
(or random energy) model with fitness values drawn from a Gaussian
distribution
\cite{OhTa90,Tach91}.
A detailed analysis  was performed by
Gillespie \cite{Gill94b,Gill95}. He observed that, under the house-of-cards
model,
populations evolve towards a state where a large
fraction of new mutations are deleterious, but of those which fix
(i.e.\ make it into the population),
one-half is advantageous, the other half is deleterious;
the reader will recognize this as a variation on the theme
`How can Muller's ratchet be halted?'.

Here, too, the mutations that fix 
are the many
ones which behave nearly neutral, that is, their fitness does not
deviate from the mean fitness by more than $1/N$; they are the relevant
ones for molecular evolution in the long run. Strongly deleterious
mutations hardly ever fix; strongly advantageous ones have become
very rare because the current well-adapted state is hard to beat
(this may be corroborated by the theory of records).

So far, everything is plausible, and one feels tempted to accept the
house-of-cards model as a model of molecular evolution. But there
is one startling surprise. Let
 $\sigma$ denote the standard deviation of the Gaussian distribution
from which the fitness values are drawn.
For $N\sigma>4$,  molecular evolution
stops \cite{Gill94b,Gill95}!  In nature,
 however,
sequence evolution proceeds for population sizes which range
over several orders of magnitude, certainly including parameter values with
$N\sigma>4$. Therefore, the house-of-cards model (with fitness values from
a Gaussian distribution) cannot be
the `right' model for molecular evolution.

Such a negative result may be a disappointing one to close with.
However, one thing is remarkable: it has been achieved to rule out
a fitness landscape on the grounds of very general observations
as well as advanced theoretical reasoning. This should be encouraging
for the theoretician.

\section{Further directions}
Quite obviously, there are more open problems than solved ones.
Let us close by emphasizing one dual pair of questions which
we think theoretical physicists could contribute to in the future.

On the theoretical side, it is imperative to
{\em explore more models.} We have seen that the
knowledge of the conditions which lead to mutational degradation
(through error thresholds, Muller's ratchet, or both)
is disappointingly sparse. All examples known so far may
be considered as case studies. It would be highly desirable
to learn about the behaviour of larger model classes,
instead of the singular examples studied so far. Which
fitness landscapes exhibit error thresholds, and which
display Muller's ratchet? How fast does the ratchet
turn? Recall that, even for the prototype model,
a full answer is not available.
More general answers to these questions could help to decide
 whether or not mutational degradation
should be considered a generic phenomenon.

On the phenomenological side, it is clear that
ultimate reasoning about fitness landscapes
must be based on  real world data. The most powerful data currently
available are sequence samples from populations, and from closely
related species. It is an important task to
 {\em improve methods for inference of evolutionary history from
sequence data.} The inference problem  we
have met in (\ref{lik}) is at the heart of this. It involves
simulation of the coalescence process, which are 
demanding even for neutral evolution. Getting them to
work under various assumptions on the fitness landscape
is a real challenge.
Computational physicists should find a lot to do here.

\subsection*{Acknowledgements}
Our views of the subject have been shaped by cooperation and fruitful
discussion with a number of physicists, in particular,
Michael Baake, Joachim Hermisson,
and Holger Wagner, who also helped with
 constructive criticism on the manuscript.
 It is our pleasure
to thank Reinhard B{\"urger}, Brian Charlesworth,
Joe Felsenstein, Wolfgang Stephan,
and G{\"u}nter Wagner for valuable discussions.

\newpage

%\bibliography{../eb}

\newcommand{\noopsort}[1]{} \newcommand{\printfirst}[2]{#1}
  \newcommand{\singleletter}[1]{#1} \newcommand{\switchargs}[2]{#2#1}

\newpage

\subsection*{Figure captions}

\noindent Figure 1:

\noindent The sampling process and the coalescent
process. In a population of constant size,
individuals are sampled with replacement
to contribute offspring to the next generation according to
Eq.~(\ref{wf_sampling}). This may be
viewed as a bifurcation process forward in time, or a
coalescent process backward in time. The genealogy of a sample
of $M=4$ alleles is indicated (fat lines).

\bigskip

\noindent Figure 2:

\noindent The genealogy of the sample in Fig.~1.
Usually, the history of the sample is unknown. Then, the genealogy
is a random variable, with the $T_m$ denoting the lengths of the
time intervals during which there are $m$ distinct lineages in the
genealogy.

\bigskip

\noindent Figure 3:

\noindent Stationary state of the paramuse model (\ref{paramuse}) and
(\ref{paramut}) for sequences of length $L=30$, and the SPL with
selective advantage $s=0.03$ of the favourable sequence. Relative
frequencies $p_j$ of sequences with $j$ `$-$' sites are represented
as a height profile over the $xy$ plane with $j$ ($0 \leq j \leq
30$) in the $x$ direction, and the mutation rate per site, $\mu$
($0 \leq \mu \leq 0.002$) in the $y$ direction. For mutation rates
above the error threshold ($\mu_c \simeq 0.001$), the population is
evenly distributed over sequence space, i.e.\ $p_j = \frac{1}{2^L}
\binom{L}{j}$.

\bigskip

\noindent Figure 4:

\noindent  Sequence genealogy as a two-dimensional, anisotropic Ising
model. Sequences descending from each other (i.e.\
grandmother-mother-daughter) form the rows of the lattice. This
way, columns correspond to sequence positions, rows to generations,
and the transition from one row to the next is governed by the
mutation-reproduction matrix ${\cal VW}$. If the letters of the
sequence are identified with spins, every genealogy corresponds to
one possible configuration of a two-dimensional Ising model, and
the transfer matrix takes the role of the mutation-reproduction
matrix. Interactions are anisotropic:  Interactions {\em between}
 the rows (dashed) are
nearest neighbour and correspond to mutation.
Interactions {\em within} the
rows (solid lines) may be arbitrary and long-range, but are
identical for every row. Their
presence or absence, as well as their strengths, define the
fitness landscape in the sense of a mapping from sequence
space into the real numbers.

\bigskip

%\noindent Figure 5:

%\noindent Stationary solutions of the decoupled mutation-selection
%model (\ref{para}) with $K=2$, $r_1=1.05$, $r_2=1$, $m_{21}=m$,
%and $m_{12}= \epsilon$, as a function of $m$. Solid lines:
%$\epsilon = 1$ (symmetric mutation), dashed: $\epsilon = 0.2$,
%dotted: $\epsilon=0$ (unidirectional mutation). The negative
%branch of the solution is biologically irrelevant and unstable,
%whereas the other one is stable. The error threshold may be
%interpreted as a transcritical bifurcation emerging from an
%avoided crossing.

%\bigskip

\noindent Figure 5:

\noindent Error thresholds in finite populations. Long-term
average of the relative frequency of the favourable sequence,
$\langle N_1/N \rangle$, as a function of the mutation rate $\mu$.
Fitness landscape: SPL with selective advantage $s=0.05$
of the fittest sequence; and sequence length:  $L=30$. Solid line:
$N=\infty$ (deterministic limiting case); dashed: $N=10000$;
dotted: $N=1000$; dash-dotted: $N=100$.

\bigskip

\noindent Figure 6:

\noindent Muller's ratchet and mutational meltdown (schematically).
Fat line: Time course of the average number of surviving offspring
per individual, $\langle R \cdot \bar w \rangle$ ($R$ is the maximum
number of offspring and $\bar w$ the mean viability). $\langle R
\cdot \bar w \rangle$ declines due to Muller's ratchet; when it falls below
1, the population goes to extinction rapidly (mutational meltdown).
Thin line: histogram of extinction probabilities. Extinction times
show surprisingly small coefficients of variation.

\newpage

\newpage

\newpage

\newpage

\newpage

%\begin{figure}
%\hspace{0cm}
%\begin{center}
%\scalebox{.75}{\includegraphics{./avcro.eps}}
%\end{center}
%\caption{\label{avcro}}
%\end{figure}

%\newpage

\newpage

\begin{figure}
\hspace{0cm}
\begin{center}
\scalebox{1.}{\includegraphics{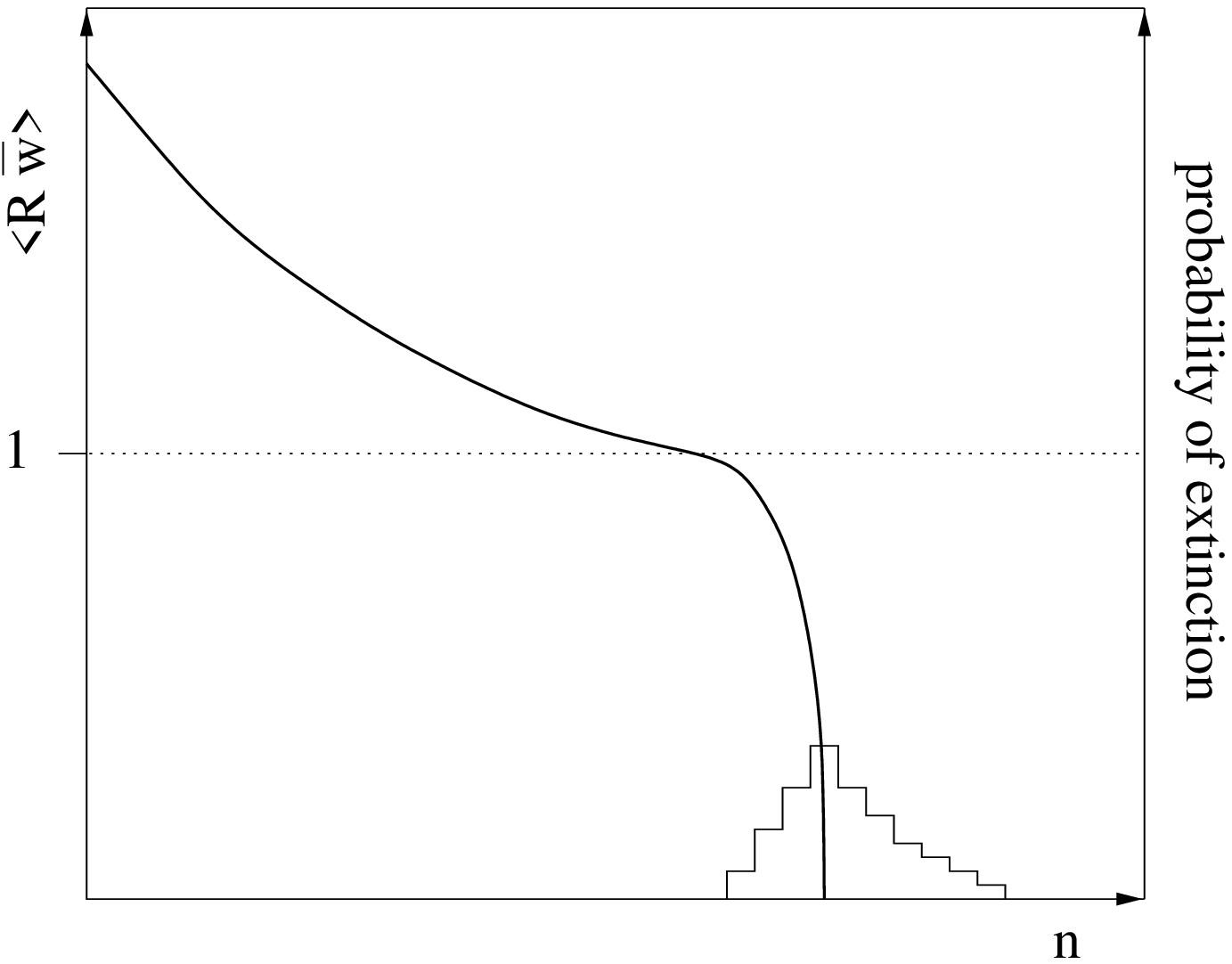}}
\end{center}
\caption{\label{meltdown}}
\end{figure}

\end{document}